\documentclass[prc,twocolumn,showpacs,preprintnumbers,amsmath,amssymb]{revtex4}
\usepackage{rotating}
\usepackage{graphicx}
\usepackage{dcolumn}
\usepackage{amstext}
\usepackage{lscape}
\usepackage{verbatim}
\usepackage[colorlinks,citecolor=blue,bookmarks]{hyperref}
\usepackage{times}
\usepackage{bm}
\usepackage{soul}
\begin{document}

\title{Unearthing radial independence for prediction of alpha-decay half-lives}
 
\author{Swagatika Bhoi*$^{1}$}
\author{Basudeb Sahu$^{2}$}

\affiliation{\it $^{1}$School of Physics, Sambalpur University, Jyoti Vihar, Burla-768019, India.} 
\affiliation{\it $^{2}$Department of Physics, College of Engineering and Technology
Bhubaneswar-751003, India.}
\date{\today}

\begin{abstract}
Nuclear radial distance is a prerequisite for generating any alpha-decay half-life formula by taking a suitable effective potential. We study the emission process of alpha particles from an isolated quasi-bound state generated by an effective potential to a scattering state. The effective potential is expressed in terms of Frahn form of potential which is exactly solvable and an analytical expression for half-life is obtained in terms of Coulomb function, wave function and the potential. We then derive a closed form expression for the decay half-life in terms of the parameters of the potential, Q-value of the system, mass and proton numbers of the nuclei valid for alpha-decay as well as proton-decay. From the nature of variations of half-life as a function of radial distance, we trace the radial independence region where decay time is almost constant. Finally by overviewing our results and picking that particular radial distance we predict the half-lives of a series of nuclei by using the closed form expression.
\end{abstract}
\maketitle
\footnotetext [1]{swagatikabhoi66@gmail.com}

\section{Introduction}\label{sec1}
The quintessential $\alpha$-radioactivity has been studied by many physicists so far and has opened doors for laying a rigid foundation and development of nuclear physics  \cite{a,b,c,d}. Gamow was the first and foremost to apply quantum mechanics to a nuclear physics problem by providing the first model to explain $\alpha$-decay and propounded that the process involves tunneling of an $\alpha$-particle through a large barrier \cite{e}. A profound knowledge of this quantum mechanical effect enables one to obtain the Geiger-Nuttall law which relates the decay constant of a radioactive isotope with the energy of the $\alpha$ particles emitted.

In other words Geiger and Nuttall were the first in giving a kickstart in the form of a simple formula for $\alpha$-decay half-lives. Following them many analytic formulas have been put forth such as Viola Seaborg-Sobiczewski (VS) formula \cite{viola,patyk,pomorski}, Ni-Ren-Dong-Xu formula \cite{ni}, Royer formula \cite{royera,royerb}, Sobiczewski Parkhomenko (SP) formula \cite{pomorski}, and Horoi formula \cite{horoi}. A few empirical relationships were also propounded such as the universal (UNIV) Curve \cite{poenaru}, the Semi-empirical formula based on fission theory (SemFIS) \cite{poenaru}, and the unified model for $\alpha$-decay and $\alpha$-capture (UMADAC) \cite{denisov,sedykh,shan}.

By considering the emission process of $\alpha$ particles in the transition from an isolated quasi-bound state to a scattering state, we have given a formula (Sahu16 formula) for $\alpha$-decay half-lives\cite{bsahu}. In another work based on the phenomena of resonances occurring in quantum scattering process under Coulomb-nuclear potential, Sahu \textit{et. al.} have derived a general decay law\cite{bsahua}. By using this general decay law along with a precise radius formula and an analytical expression for preformation probability, an improved semi-empirical relationship (ImSahu formula) for $\alpha$-decay half-lives has been proposed and thereby the accuracy has been improved significantly \cite{shan}.

It is customary to mention here that it is not our motive to perturb the Sahu16 formula. Infact, we intend to find the half-lives by introducing another form of potential markedly different from the potential used in \cite{bsahu}. From the decay half-lives we are interested in tracking the radial independence which will help us in ascertaining the radial distance expression to be used to predict the half-lives by Sahu16 formula. Thus by following the same formalism and applying the Sahu16 formula, we stress on its applicability to alpha decay as well as proton decay. 

The paper is organized as follows: In the next section we will mention the decay half-life formula.  Also we will present the closed form expression. In 
section \ref{sec3} we will move on to the applicability of the Sahu16 formula. 
In section \ref{conc} we finally summarize the paper. 

\section{Theoretical framework}\label{sec2}
\subsection{Decay width or half-life of $\alpha$-decay}
We remark that in the $\alpha$-decay process the $\alpha$-cluster in the decaying nucleus is controlled by an attractive nuclear potential, $V_N(r)$ and the $\alpha$-particle outside the nucleus by the point-charge Coulomb potential given by $V_C^P=\frac{Z_{\alpha}Z_De^2}{r}$,
where $Z_{\alpha}$ and $Z_D$ are the proton numbers of the $\alpha$-particle and daughter nucleus respectively and $e^2$=1.43996 MeV fm.

In a simple picture we represent $H-H_0$ as the difference between the potentials in the two cases viz. the nuclear+Coulomb potential and the point-charge Coulomb potential i.e
\begin{equation}
H-H_0=\lbrace V_N(r)+V_C(r)\rbrace-V_C^P(r)=V_{eff}(r)-V_C^P(r),
\end{equation}
where $V_{eff}(r)$ is the effective potential and $V_C(r)$ is the Coulomb potential given by 
\begin{equation}
V_C(r)=
\left\{
\begin{array}{cl}
\frac{Z_{\alpha}Z_De^2}{2R_C}\lbrack3-(\frac{r}{R_C})^2\rbrack
&{if\;\;\;\;\; 0<r<R_C,}\\
\frac{Z_{\alpha}Z_De^2}{r} &{if\;\;\;\;\; r>R_C,}
\end{array}
\right .
\end{equation}
where, $R_C$ is the Coulomb radius parameter having value $R_C=r_c(A_\alpha^{1/3}+A_D^{1/3})$, $r_c=1.2$ fm which is the distance parameter. $A_\alpha$ represent the mass number of $\alpha$ particle, $A_D$ represent the mass number of the daughter nucleus.
In our approach, we calculate the decay width by taking into account the $\alpha$-decay process where there is transition of an $\alpha$-cluster from an isolated quasi-bound state to a scattering state. The initial system is related with the instability with the quasi-bound state of the decaying nucleus. Along with that the final state is the scattering state of the $\alpha$+daughter system.

Now, we solve the Schr\"{o}dinger equation using the effective potential which is the amalgamation of the nuclear potential $V_N(r)$ and the electrostatic potential $V_C(r)$ to get the radial part of the initial and final state of the wave function.

The radial part of the initial state wave function is
\begin{equation}
\psi_{nl}(r)=\frac{u_{nl}(r)}{r}.
\end{equation}

Additionally, the final state wave function can be written considering the motion of the $\alpha$-particle relative to the daughter nucleus as a scattering state wave function corresponding to the $\alpha$-particle in point charge Coulomb potential \cite{fur}:
\begin{equation}
\phi(r)=\sqrt \frac{2\mu}{\pi \hbar^2 k}\hspace{2mm}\frac{F_l(r)}{r},
\end{equation}
where $k=\sqrt{2\mu E_{c.m.}/\hbar}$, $E_{c.m.}$ stands for the center-of-mass energy, $\mu=m_n\frac{A_{\alpha}A_D}{A_{\alpha}+A_D}$ is the reduced mass of the system with $m_n$ giving the mass of a nucleon and $F_l(r)$ is the regular Coulomb wave function for a given partial wave $l$. The factor $\sqrt{\frac{2\mu}{\pi \hbar^2k}}$ is a normalization factor of the scattering wave function.

The effective potential as a function of distance can be solved exactly for the wave function $u_{nl}(r)$ for $l=0$ which is covered elaborately in the upcoming section.
Based on the Gell-Mann-Goldberger transformation \cite{dav}, the expression for the decay width becomes
\begin{equation}
\Gamma=\frac{4\mu}{\hbar^2k}\hspace{2mm}\frac{|\int_0^RF_l(r)\;\lbrack V_{eff}(r)-V_C^P(r)\rbrack u_{nl}(r)dr|^2}{\int_0^R|u_{nl}(r)|^2dr}.
\end{equation}
For the normalization of the interior wave function the factor $\int_0^R|u_{nl}(r)|^2dr$ is used. The resonant wave function $u_{nl}(r)$ decreases rapidly with distance outside the Coulomb barrier radius $R_0$. For this reason, we apply the box normalization condition for the wave function $\int_0^R|u_{nl}(r)|^2dr=1$ for $R\approx R_0$. So, the preformation probability here is taken as 1 as the particle has been formed already. 
We are quite familiar with the relation between decay half-life $T_{1/2}$ and the width:
\begin{equation}
T_{1/2}=\frac{\hbar\; ln2}{\Gamma}.
\end{equation}
By using (5), we get a new expression of $T_{1/2}$
\begin{equation}
T_{1/2}=\frac{0.693\hbar^3k}{4\mu}\hspace{2mm}\frac{1}{J},
\end{equation}
\begin{equation}
J=\vert\int_0^RF_l(r)\;\lbrack V_{eff}(r)-V_C^P(r)\rbrack u_{nl}(r)dr\vert^2.
\end{equation}
Now, the regular Coulomb wave function $F_l(r)$ can be expressed as \cite{fro} 
\begin{equation}
F_l(r)=A_l\rho^{l+1}f_l(\rho),
\end{equation}
where $\rho=kr$, Sommerfeld parameter $\eta=\frac{\mu}{\hbar^2}\hspace{2mm}\frac{Z_{\alpha}Z_De^2}{k}$,
\begin{equation}
f_l(\rho)=\int_0^\infty (1-\tanh^2\epsilon)^{l+1}\hspace{1mm}\cos(\rho \tanh\epsilon-2\eta\epsilon)\hspace{1mm}d\epsilon,
\end{equation}
\begin{equation}
A_l=\frac{\sqrt {1-\exp(-2\pi\eta)}}{2^l\lbrace 2\pi\eta(1+\eta^2)(2^2+\eta^2)\ldots(l^2+\eta^2)\rbrace^{1/2}}.
\end{equation}
In particular, for $l=0$, $A_l$ is given by
\begin{equation}
A_0=\left\{\frac{1-\exp(-2\pi\eta)}{2\pi\eta}\right\}^{\frac{1}{2}}.
\end{equation}

\subsection{The effective $\alpha$+nucleus potential}
The  sum of the nuclear potential $V_N(r)$ and the Coulomb potential $V_C(r)$ i.e. the effective potential mentioned in (eqn.1) can be represented in the Frahn form of potential given by \cite{fie}:
\begin{equation}
V_{eff}(r)=
\left\{
\begin{array}{cl}
V_0[S_1+(S_2-S_1)\rho _1]
&{if\;\;\;\;\; r \le R_0,}\\
V_0S_2\rho _2 &{if\;\;\;\;\; r \ge R_0,}
\end{array}
\right .
\end{equation}
where $V_0$ is the strength of the potential with value $1$ MeV.
\begin{equation}
	\rho_n=\frac{1}{\cosh^2(\frac{R_0-r}{d_n})};n=1,2,\nonumber\\
\end{equation}
$d_n$ accounts for the flatness of the barrier, $d_1$ deciding the steepness of the interior side of the barrier whereas the exterior side is judged by $d_2$. $R_0$ is the barrier radius having value; $R_{0}=r_{0}(A_\alpha ^{1/3}+A_D^{1/3})+2.72$, $r_{0}=0.97$ fm. Moreover $S_1$ and $S_2$ are the depth and height of the potential, respectively, having values;
\begin{equation}
S_1=-78.75+\frac{3Z_\alpha Z_D e^2}{2R_c},\nonumber \\ 
\end{equation}
\begin{equation}
S_2=\frac{Z_\alpha Z_D e^2}{R_0}(1-\frac{a_g}{R_0}),\nonumber\\
\end{equation}
where $a_g=1.6$ fm is the distance parameter.

The potential given by (eqn.13) can generate resonance states having a pocket inside and a barrier outside. This potential in the Schrodinger equation is solved exactly for the eigen function.
 
The reduced S-wave Schr\"{o}dinger equation for the region $r\le R_0$ is written in the dimensionless form as follows:
\begin{eqnarray}
\frac{d^2u_1}{dr^2}+[\kappa^2-k_{0}^{2}S\rho_1]u_1=0,
\end{eqnarray}
where $\kappa^2=k^2-k_{0}^{2} S_1$, $\frac{2\mu}{\hbar^2} V_0=k_0^2$, $S=S_2-S_1$.

The solution $u_{1}(r)$ in the region $r\le R_0$ is given by
\begin{equation}
u_1(r)=A_1 z_1^{{i/2}\kappa d_1} F(a_1,b_1,c_1,z_1) 
 + B_1 z_1^{-i/2\kappa d_1} F(a_1^\prime,b_1^\prime,c_1^\prime,z_1^\prime),\\
\end{equation}
\begin{equation}
a_{1}=\frac{1}{2}({\lambda}_{1}+i{\kappa}d_{1}),b_{1}=\frac{1}{2}({1-\lambda}_{1}+i{\kappa}d_{1}),c_{1}=1+i{\kappa}d_{1},\\
\end{equation}
\begin{equation}
a_{1}^{\prime}=\frac{1}{2}({\lambda}_{1}-i{\kappa}d_{1}),b_{1}^{\prime}=\frac{1}{2}({1-\lambda}_{1}-i{\kappa}d_{1}),c_{1}^{\prime}=1-i{\kappa}d_{1},\\
\end{equation}
\begin{equation}
{\lambda}_1=\frac{1}{2}-\frac{1}{2}[1-(2{\kappa}d_1)^2S]^{1/2},\\
\end{equation}
where $z_1=\rho_1(r)$ and $F(a,b,c,z)$ is the hypergeometric function. 

Using the boundary condition $u_1\rightarrow0$ for $r\rightarrow0$, we get
\begin{equation}
\rho_1(r=0)=z_0, \rho_1(r=0)=\frac{1}{cosh^2(\frac{R_0}{d_1})},
\end{equation}
Moreover, as $R_0>>d_1$, $cosh^2(\frac{R_0}{d_1})>>1$ and $\rho_1(r=0)=z_0<<1$ we can write
\begin{equation}
z_0=\frac{1}{R_0/d_1}=\frac{4}{(e^{R_0/d_1}+e^{-R_0/d_2})^2}\\
\end{equation}
\begin{equation}
\Rightarrow z_0=2^2e^{-R_0/d_1} (Since, e^{-R_0/d_1}<<1 )\\
\end{equation}
Thus,
\begin{equation}
C_1=-\frac{B_1}{A_1}=z_0^{i\kappa d_1}\frac{F(a_1,b_1,c_1,z_0)}{F(a_1^{\prime},b_1^{\prime},c_1^{\prime},z_0)},
\end{equation}
\begin{equation}
\simeq z_0^{i\kappa d_1}, (Since, F(a,b,c,0)=1)\nonumber\\
\end{equation}
\begin{equation}
\Rightarrow C_1=e^{-2i\kappa R_0+2i\kappa d_1ln2}=e^{-2i\kappa(R_0-d_1ln2)}.\\
\end{equation}

\begin{figure}
\includegraphics[width=1.1\columnwidth,clip=true]{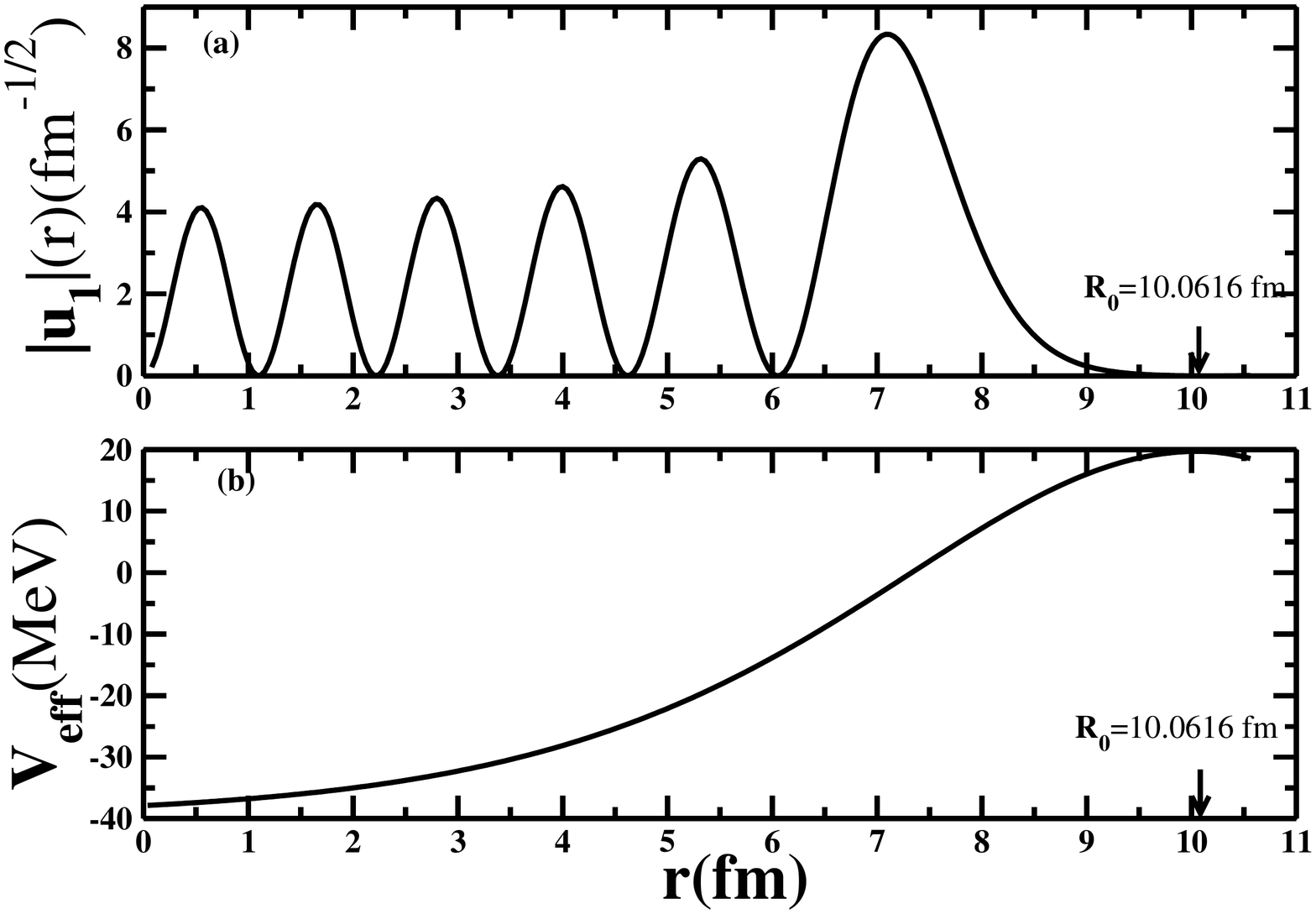}
\vspace{0.2cm} \caption{Various terms for explaining the $\alpha$-decay rate in S-wave of $\alpha+_{82}^{214}$Pb system having barrier radius $R_0$=10.0616 fm : (a) the modulous of the radial wave function $|u_1(r)|$ at resonance, (b) the $\alpha+$ daughter potential $V_{eff}$ as a function of radial distance r in fm.}\label{figOne.}
\label{fig1}
\end{figure}

We make a lay out (Figs. \ref{fig1}(a) and (b)) to properly describe the modulous of the resonance state wave function, $|u_1(r)|$ and the combined nuclear and Coulomb potential, $V_{eff}(r)$ by taking the $\alpha$+daughter system ($\alpha+\hspace{0.5mm}_{82}^{214}$Pb) with Q-value of decay or energy $E=6.115$ MeV. It is visible from the plot \ref{fig1}(a) that the wave function of the resonance state decreases exponentially in the barrier region having Coulomb barrier radius, $R_0=10.0616$ fm. We are very much concerned with this particular region since the diminishing of the wave function is observed here.

By taking the above found solution $u_1(r)$ as the wave function $u_{nl}(r)$ for $l=0$, Coulomb function $F_l(r)$ and the effective potential $V_{eff}(r)$ in the formula for decay width i.e. $\Gamma$ (eqn.5) and using the $\Gamma$ in (eqn.6), we calculate the half-life and represent it as $T_{1/2}^{calt.}$.

\subsection{Closed form expression for decay half-life}

We now consider the problem of $\alpha+$nucleus system with a specific energy value $Q_{\alpha}$, the mentioned radius $R=R_0$, the value of Sommerfeld parameter $\eta$ and parameter $\rho=kR$ are such that $\eta\rho\le50$ and $\rho\approx 10$. In this context, we use the same power series expansion and the expression for Coulomb wave function $F_l^{ps}(r)$ as done in \cite{bsahu}.
\begin{equation}
F_l^{ps}(r)=C_l\rho^{l+1}G_l,
\end{equation}
\begin{equation}
(n+1)(n+2l+2)G_{n+1}=2\eta\rho G_n-\rho^2G_{n-1},
\end{equation}
\begin{equation}
G_0=1, G_1=\frac{\eta\rho}{(l+1)}, G_l=\sum_{j=1}^{500} G_j,
\end{equation}
\begin{equation}
C_l^2=\frac{P_l(\eta)}{2\eta}\frac{C_0^2(\eta)}{(2l+1)},
\end{equation}
\begin{equation}
P_l(\eta)=\frac{2\eta(1+\eta^2)(4+\eta^2)\ldots(l^2+\eta^2)2^{2l}}{(2l+1)\lbrack(2l)!\rbrack^2}.
\end{equation}

We find that 
\begin{equation}
F_l(r)=x_mF_l^{ps}(r),
\end{equation}
where $x_m\approx70$ for $\alpha$-decay and $x_m\approx5.9$ for proton decay. Therefore, instead of computing function $F_l(r)$ using (eqn.9), we go by the simple power series expansion of $F_l^{ps}$ multiplied by a factor $x_m=70$ for $\alpha$-decay and $x_m=5.9$ for proton decay. The magnitude of this function is zero near the origin $r=0$ but increases predominantly at $r=R_0$ whereas the resonant wave function $u_1(r)$ is very small beyond $r=R_0$. Hence, the integral $J$ (eqn.8) can be written in terms of $F_l(r)=x_mF_l^{ps}(r)$ at a point $r=R=R_0$ along with some multiplying factor which take care of the other contributions within the region $0<r<R$.

The integral $J$ now changes to
\begin{equation}
J=|c_fF_l(R)|^2=|c_fx_mF_l^{ps}(R)|^2,
\end{equation}
The value of $c_f$ can now be written as
\begin{equation}
c_f=\frac{\sqrt J}{|x_m F_l^{ps}(R)|}.
\end{equation}

The decay half-life $T_{1/2}$ in logarithmic form  can be written as

\begin{equation}
	log(T_{1/2})=a\chi+c+d+b_l,
\end{equation}
where the parameters,
\begin{equation}
a=0.9889,
\end{equation}
\begin{equation}
\chi=Z_{\alpha}Z_D\sqrt{\frac{A_{\alpha}A_D}{(A_{\alpha}+A_D)Q_{\alpha}}},
\end{equation}
\begin{equation}
c=-2logS; S=c_fx_mRG_l\frac{A_{\alpha}A_D\sqrt{Z_{\alpha}Z_D}}{A_{\alpha}+A_D},
\end{equation}

\begin{equation}
d=-45.2631,
\end{equation}

\begin{equation}
b_l=log(q_l), q_l=\frac{2\eta(2l+1)}{P_l(\eta)\rho^{2l}}.
\end{equation}

The expression (eqn.32) is some what similar to the Viola-Seaborg relation \cite{viola,tieukuang} but the difference is that in the present case the parameters and coefficients namely $a$, $c$, $d$, $b_l$ are well defined \cite{bsahu}.

\begin{table*}
\hspace{0.2 cm}
	\caption{The $\alpha$-decay energies $Q_{\alpha}$ in MeV and the experimental results of half-lives $log_{10}(T_{1/2}^{expt.})$=$\tau^{expt.}$ in seconds \cite{shan}. The value of the parameter $d_1$ representing the inner steepness of the barrier and $d_2$ is kept constant throughout i.e. $d_2=2$. Logarithm of calculated $\alpha$-decay half-lives 
$log_{10}(T_{1/2}^{calt.})$=$\tau^{calt.}$ in seconds using (eqn.7), logarithm of predicted $\alpha$-decay half-lives $log_{10}(T_{1/2}^{pred.})$=$\tau^{pred.}$ in seconds using (eqn.32) with parameter fixed $c_f=0.19$, $r_0=0.98$ for $l=0$. $R_0$ is the barrier radius in fm and $R_w$ is the radial independence region in fm.}
\renewcommand{\tabcolsep}{.4cm}
\renewcommand{\arraystretch}{1.5}
{\begin{tabular}{|c|c|c|c|c|c|c|c|c|c|c|c|c|c|c|c|c|}
\hline
\cline{2-8}
\hline
Nucleus &       Q$_\alpha$ (MeV)  &  $d_1$&     $\tau^{expt.}$ (s) &  $\tau^{calt.}$ (s)&$\tau^{pred.}$ (s)&$R_0$(fm)  &   $R_w$(fm) \\
\hline
$^{  106}_{   52}$&       4.290&    6.4152&      -4.222&      -4.560 &	-3.886	& 8.7918  &  8.9-21.0\\
$^{  112}_{   54}$&       3.330&    6.2279&       2.528&       2.199 &	2.600	& 8.8790  &  8.9-21.3\\
$^{  114}_{   56}$&       3.534&    4.7248&       1.679&       1.565 &	2.515	& 8.9073  &  8.9-21.5\\
$^{  146}_{   62}$&       2.528&    4.9078&       15.512&      15.152 &15.574	& 9.3202  &  8.9-21.1\\
$^{  148}_{   64}$&       3.271&    6.3576&       9.372&       9.700 &	9.414	& 9.3439  &  9.2-15.0\\
$^{  150}_{   66}$&       4.351&    6.4719&       3.115&       3.311 &	3.103	& 9.3673  &  9.1-15.0\\
$^{  154}_{   70}$&       5.474&    6.4903&      -0.355&      -0.114 &	 -0.332	& 9.4135  &  9.3-15.7\\
$^{  156}_{   72}$&       6.028&    6.4981&      -1.638&      -1.391 &	-1.613	& 9.4363  &  9.2-15.2\\
$^{  162}_{   76}$&       6.767&    5.0320&      -2.688&      -2.957 &	 -2.525	& 9.5036  &  9.5-15.6\\
$^{  168}_{   78}$&       6.999&    5.0616&      -2.696&      -2.886 &	 -2.524	& 9.5691  &  9.5-15.5\\
$^{  190}_{   84}$&       7.699&    5.2129&      -2.595&      -2.729 &	 -2.579	& 9.7966  &  9.6-15.9\\
$^{  218}_{   84}$&       6.115&    4.1634&       2.269&       2.229 &	 2.760	& 10.0616 &  9.6-15.1\\
$^{  194}_{   86}$&       7.862&    3.9172&      -3.108&      -3.158 &	 -2.340	& 9.8361  &  9.5-15.3\\ 
$^{  218}_{   86}$&       7.263&    4.2233&      -1.455&      -1.364 &	  -0.757& 10.0616 &  10.0-15.0\\
$^{  220}_{   86}$&       6.405&    4.1213&       1.746&       1.911 &	 2.425	& 10.0796 &  10.0-15.0\\
$^{  224}_{   88}$&       5.789&    3.9714&       5.523&       5.707 &	 6.118	& 10.1153 &  10.1-14.7\\
$^{  216}_{   90}$&       8.071&    5.3465&      -1.583&      -1.583 &	-1.716	& 10.0435 &  10.1-14.3\\
$^{  218}_{   90}$&       9.849&    5.7190&      -6.932&      -6.339 &	-6.250	& 10.0616 &  10.1-14.3\\
$^{  224}_{   92}$&       8.633&    4.1746&      -3.387&      -3.279 &	-2.686	& 10.1153 &  10.1-14.7\\
$^{  232}_{   94}$&       6.716&    2.7289&       4.122&       3.409 &	 4.564	& 10.1855 &  10.2-14.2\\ 

\hline
\end{tabular}}
\label{tab1}
\end{table*}

\begin{table*}
\hspace{0.2 cm}
	\caption{The proton-decay energies $Q_{p}$ in MeV and the experimental results of half-lives $log_{10}(T_{1/2}^{expt.})$=$\tau^{expt.}$ in seconds \cite{bsahua}. The value of the parameter $d_1$ representing the inner steepness of the barrier and $d_2$ is kept constant throughout i.e. $d_2=2$. Logarithm of calculated proton-decay half-lives $log_{10}(T_{1/2}^{calt.})$=$\tau^{calt.}$ in seconds using (eqn.7), logarithm of predicted proton-decay half-lives 
$log_{10}(T_{1/2}^{pred.})$=$\tau^{pred.}$ in seconds using (eqn.32) with parameter fixed $c_f=2.8$, $r_0=0.97$ for $l=0$. $R_0$ is the barrier radius in fm and $R_w$ is the radial independence region in fm.}
\renewcommand{\tabcolsep}{.4cm}
\renewcommand{\arraystretch}{1.5}
{\begin{tabular}{|c|c|c|c|c|c|c|c|c|c|c|c|c|c|c|c|c|}
\hline
\cline{2-8}
\hline
Nucleus &       Q$_p$ (MeV)  &  $d_1$&     $\tau^{expt.}$ (s) &  $\tau^{calt.}$ (s)&$\tau^{pred.}$ (s)&$R_0$(fm)&   $R_w$(fm) \\
\hline
$^{  157}$Ta&        0.947&    4.5600&       -0.523&        -0.389 &	0.218	& 8.9116 &  8.5-10.5\\
$^{  167}$Ir&       1.086&    4.6329&       -5.180&       -1.566 &	-0.970	& 9.0208 &  8.6-10.5\\
$^{  185}$Bi&       1.624&    4.8270&       0.523&       -5.092  &	-5.058	& 9.2070 &  8.7-10.6\\
$^{  171}$Au&       1.469&    4.6970&       -0.959&      -4.845 &     -4.741	& 9.0633 &  8.4-10.7\\
$^{  177}$Tl&       1.180&    4.6939&       -0.959&      -1.021 &      -0.930	& 9.1258 &  8.3-10.1\\


\hline
\end{tabular}}
\label{tab2}
\end{table*}

\begin{table*}
\hspace{0.2 cm}
	\caption{ List of $\alpha$-decay half-lives for 144 e-e nuclei. Logarithm of predicted $\alpha$-decay half-lives $log_{10}(T_{1/2}^{pred.})$=$\tau^{pred.}$ in seconds using (eqn.32) with parameter fixed $c_f=0.19$, $r_0=0.98$ for $l=0$. The experimental results of half-lives $log_{10}(T_{1/2}^{expt.})$= $\tau^{expt.}$ in seconds \cite{shan}. The $\alpha$-decay energies $Q_{\alpha}$ are in MeV.}
\renewcommand{\tabcolsep}{.1 cm}
\renewcommand{\arraystretch}{1.3}
{\begin{tabular}{|c|c|c|c|c|c|c|c|c|c|c|c|c|c|}
\hline
\cline{2-12}
\hline
Nucleus &       Q$_\alpha$ (MeV)  &      $\tau^{expt.}$ (s) &  $\tau^{pred.}$ (s)&Nucleus &       Q$_\alpha$ (MeV)  &      $\tau^{expt.}$ (s) &  $\tau^{pred.}$ (s)&Nucleus &       Q$_\alpha$(MeV) &        $\tau^{expt.}$ (s) &  $\tau^{pred.}$ (s)\\
\hline
 $^{  106}_{   52}$&       4.290&          -4.222&      -3.886	&	$^{  108}_{   52}$&       3.445&            0.632&        0.569&
 $^{  112}_{   54}$&       3.330&           2.528&       2.600\\
 $^{  114}_{   56}$&       3.534&           1.679&       2.515	&	$^{  146}_{   62}$&       2.528&          15.512&      15.574&
$^{  148}_{   62}$&       1.986&          23.344&      23.604\\
 $^{  148}_{   64}$&       3.271&           9.372&       9.414	&	$^{  150}_{   64}$&       2.809&          13.752&      13.859& 
 $^{  152}_{   64}$&       2.204&          21.533&      21.757\\
 $^{  150}_{   66}$&       4.351&           3.115&       3.103	&	$^{  152}_{   66}$&       3.726&           6.933&       7.100& 
$^{  154}_{   66}$&       2.945&          13.976&      13.880\\
 $^{  152}_{   68}$&       4.934&           1.054&       1.118	&	$^{  154}_{   68}$&       4.280&           4.678&       4.626&
$^{  154}_{   70}$&       5.474&           -.355&       -0.332\\
 $^{  156}_{   70}$&       4.811&           2.417&       2.748	&	$^{  158}_{   70}$&       4.172&           6.629&       6.433&
$^{  156}_{   72}$&       6.028&          -1.638&      -1.613\\
 $^{  158}_{   72}$&       5.405&            0.808&        0.903	&	$^{  160}_{   72}$&       4.902&           3.288&       3.297&
$^{  158}_{   74}$&       6.613&          -2.903&      -2.813\\
 $^{  160}_{   74}$&       6.066&           -0.980&       -0.893	&	 $^{  162}_{   74}$&       5.674&            0.478&        0.658&
$^{  166}_{   74}$&       4.856&           4.739&       4.536\\
 $^{  162}_{   76}$&       6.767&          -2.688&      -2.525	&	$^{  166}_{   76}$&       6.139&           -0.539&       -0.340&
$^{  170}_{   76}$&       5.539&           1.904&       2.119\\
 $^{  174}_{   76}$&       4.872&           5.342&       5.423	&	$^{  168}_{   78}$&       6.999&          -2.696&      -2.524&
$^{  170}_{   78}$&       6.708&          -1.847&      -1.584\\
 $^{  174}_{   78}$&       6.184&            0.073&        0.288	&	$^{  176}_{   78}$&       5.885&           1.197&       1.483&
$^{  178}_{   78}$&       5.573&           2.722&       2.841\\
 $^{  180}_{   78}$&       5.240&           4.301&       4.436	&	$^{  182}_{   78}$&       4.952&           5.577&       5.949&
$^{  184}_{   78}$&       4.598&          7.812&       8.020\\
 $^{  188}_{   78}$&       4.008&          12.547&      12.091	&	$^{  190}_{   78}$&       3.249&          19.312&      19.008&
$^{  174}_{   80}$&       7.233&          -2.720&      -2.528\\
$^{  176}_{   80}$&       6.908&          -1.639&      -1.498	&	$^{  180}_{   80}$&       6.258&            0.731&        0.827&
$^{  182}_{   80}$&       5.997&           1.859&       1.870\\
 $^{  184}_{   80}$&       5.662&          3.393&       3.332	&	$^{  188}_{   80}$&       4.705&           8.734&       8.428&
$^{  186}_{   82}$&       6.470&           1.081&        0.817\\
 $^{  188}_{   82}$&       6.109&           2.436&       2.251	&	$^{  190}_{   82}$&       5.698&           4.249&       4.064&
 $^{  210}_{   82}$&       3.792&          16.567&      16.164\\
 $^{  190}_{   84}$&       7.699&          -2.595&      -2.579	&	$^{  192}_{   84}$&       7.320&          -1.480&      -1.423&
 $^{  194}_{   84}$&       6.987&           -0.375&       -0.326\\
 $^{  196}_{   84}$&       6.658&            0.775&        0.847	&	$^{  198}_{   84}$&       6.309&           2.270&       2.201&
$^{  200}_{   84}$&       5.981&           3.794&       3.589\\
 $^{  202}_{   84}$&       5.701&           5.144&       4.872	&	$^{  204}_{   84}$&       5.485&           6.285&       5.927&
$^{  206}_{   84}$&       5.237&           7.145&       7.228\\
 $^{  208}_{   84}$&       5.215&           7.961&       7.324	&	$^{  210}_{   84}$&       5.407&           7.708&       6.254&
$^{  212}_{   84}$&       8.954&          -6.524&      -6.173\\
 $^{  214}_{   84}$&       7.834&          -3.786&      -3.290	&	$^{  216}_{   84}$&       6.906&           -0.839&       -0.328&
$^{  218}_{   84}$&       6.115&           2.269&       2.760\\
 $^{  194}_{   86}$&       7.862&          -3.108&      -2.340	&	$^{  196}_{   86}$&       7.617&          -2.356&      -1.620&
$^{  198}_{   86}$&       7.349&          -1.184&       -0.783\\
 $^{  200}_{   86}$&       7.043&           -0.009&        0.241	&	$^{  202}_{   86}$&       6.774&           1.095&       1.201&
 $^{  206}_{   86}$&       6.384&           2.732&       2.696\\
 $^{  208}_{   86}$&       6.261&           3.372&       3.189	&	$^{  210}_{   86}$&       6.159&           3.954&       3.606&
$^{  212}_{   86}$&       6.385&           3.157&       2.612\\
 $^{  218}_{   86}$&       7.263&          -1.455&       -0.757	&	$^{  220}_{   86}$&       6.405&           1.746&       2.425&
 $^{  222}_{   86}$&       5.590&           5.519&       6.158\\
 $^{  204}_{   88}$&       7.636&          -1.244&       -0.992	&	$^{  206}_{   88}$&       7.415&           -0.620&       -0.294&
$^{  208}_{   88}$&       7.273&            0.136&        0.163\\
 $^{  210}_{   88}$&       7.152&            0.568&        0.562	&	$^{  212}_{   88}$&       7.032&           1.185&        0.968&
 $^{  214}_{   88}$&       7.273&            0.392&        0.084\\
 $^{  220}_{   88}$&       7.592&          -1.740&      -1.059	&	$^{  222}_{   88}$&       6.679&           1.572&       2.183&
$^{  224}_{   88}$&       5.789&           5.523&       6.118\\
 $^{  226}_{   88}$&       4.871&          10.731&      11.336	&	$^{  210}_{   90}$&       8.053&          -2.046&      -1.583&
 $^{  216}_{   90}$&       8.071&          -1.583&      -1.716\\
 $^{  218}_{   90}$&       9.849&          -6.932&      -6.250	&	 $^{  222}_{   90}$&       8.127&          -2.640&      -1.960&
$^{  224}_{   90}$&       7.298&            0.124&        0.695\\
 $^{  226}_{   90}$&       6.451&           3.385&       3.971	&	$^{  228}_{   90}$&       5.520&          7.915&       8.480&
$^{  230}_{   90}$&       4.770&          12.494&      13.088\\
 $^{  232}_{   90}$&       4.082&          17.752&      18.441	&	$^{  224}_{   92}$&       8.633&          -3.387&      -2.686&
$^{  226}_{   92}$&       7.715&           -0.385&        0.057\\
 $^{  228}_{   92}$&       6.803&           2.905&       3.370	&	$^{  230}_{   92}$&       5.993&           6.426&       6.975&
$^{  232}_{   92}$&       5.414&           9.504&      10.060\\
 $^{  234}_{   92}$&       4.860&          13.036&      13.543	&	$^{  236}_{   92}$&       4.573&          14.999&      15.592&
 $^{  238}_{   92}$&       4.270&          17.252&      17.987\\
 $^{  230}_{   94}$&       7.180&           2.100&       2.726	&	$^{  232}_{   94}$&       6.716&           4.122&       4.564&
 $^{  234}_{   94}$&       6.310&           5.888&       6.346\\
$^{  236}_{   94}$&       5.867&           8.116&       8.517	&	$^{  238}_{   94}$&       5.593&           9.591&       9.985&
 $^{  240}_{   94}$&       5.256&          11.454&      11.963\\
 $^{  242}_{   94}$&       4.985&          13.187&      13.700	&	$^{  244}_{   94}$&       4.666&          15.502&      15.952&
 $^{  238}_{   96}$&       6.620&           5.510&       5.802\\
 $^{  240}_{   96}$&       6.398&           6.517&       6.788	&	$^{  242}_{   96}$&       6.216&           7.278&       7.635&
$^{  244}_{   96}$&       5.902&           8.871&       9.214\\
 $^{  246}_{   96}$&       5.475&          11.262&      11.599	&	$^{  248}_{   96}$&       5.162&          13.166&      13.537&
$^{  240}_{   98}$&       7.719&           1.990&       2.272\\
 $^{  244}_{   98}$&       7.329&           3.342&       3.664	&	 $^{  246}_{   98}$&       6.862&           5.210&       5.538&
$^{  248}_{   98}$&       6.361&           7.557&       7.794\\
 $^{  250}_{   98}$&       6.128&           8.699&       8.929	&	 $^{  252}_{   98}$&       6.217&           8.010&       8.452&
$^{  254}_{   98}$&       5.926&           9.308&       9.942\\
 $^{  246}_{  100}$&       8.374&            0.172&        0.780	&	 $^{  248}_{  100}$&       8.002&           1.687&       1.982&
$^{  250}_{  100}$&       7.556&           3.380&       3.557\\
 $^{  252}_{  100}$&       7.153&           5.037&       5.113	&	 $^{  254}_{  100}$&       7.307&           4.138&       4.467&
$^{  256}_{  100}$&       7.027&           5.137&       5.587\\
 $^{  252}_{  102}$&       8.548&            0.680&        0.910	&	 $^{  256}_{  102}$&       8.581&            0.526&        0.757&
$^{  256}_{  104}$&       8.926&            0.319&        0.441\\
 $^{  258}_{  104}$&       9.190&          -1.035&       -0.356	&	 $^{  260}_{  106}$&       9.901&          -1.686&      -1.618&
$^{  262}_{  106}$&       9.600&          -1.504&       -0.846\\
 $^{  264}_{  108}$&      10.591&          -2.796&      -2.712	&	 $^{  266}_{  108}$&      10.346&          -2.638&      -2.148&
 $^{  270}_{  108}$&       9.050&            0.556&       1.349\\
 $^{  270}_{  110}$&      11.120&          -4.000&      -3.366	&	 $^{  286}_{  114}$&      10.350&           -0.699&       -0.432&
 $^{  288}_{  114}$&      10.072&          -0.180&        0.285\\
 $^{  290}_{  116}$&      10.990&          -1.824&      -1.437	&	 $^{  292}_{  116}$&      10.774&          -1.745&       -0.932&
 $^{  294}_{  118}$&      11.820&          -3.161&      -2.767\\


\hline
\end{tabular} }
\label{tab3}
\end{table*}

\begin{table*}
\hspace{0.2 cm}
	\caption{List of $\alpha$-decay half-lives for 112 e-o nuclei. Logarithm of predicted $\alpha$-decay half-lives $log_{10}(T_{1/2}^{pred.})$=$\tau^{pred.}$ in seconds using (eqn.32) with parameter fixed $c_f=0.19$, $r_0=0.98$ for $l=0$, $c_f=0.02$ for $l>0$. The experimental results of half-lives 
$log_{10}(T_{1/2}^{expt.})$=$\tau^{expt.}$ in seconds \cite{shan}. The $\alpha$-decay energies $Q_{\alpha}$ are in MeV.}
\renewcommand{\tabcolsep}{.1 cm}
\renewcommand{\arraystretch}{1.3}
{\begin{tabular}{|c|c|c|c|c|c|c|c|c|c|c|c|c|c|c|c|c|}
\hline
\cline{2-14}
\hline
Nucleus &       Q$_\alpha$ (MeV)  &  $l$&     $\tau^{expt.}$ (s) &  $\tau^{pred.}$ (s)&Nucleus &       Q$_\alpha$ (MeV)  &$l$&       $\tau^{expt.}$ (s) &  $\tau^{pred.}$ (s)&Nucleus &       Q$_\alpha$ (MeV) &$l$&        $\tau^{expt.}$ (s) &  $\tau^{pred.}$ (s)\\
\hline
 $^{  105}_{   52}$&       4.889&    0&      -6.208&      -6.277& $^{  107}_{   52}$&       4.008&    0&      -2.354&      -2.572&$^{  109}_{   54}$&       4.217&    2&      -1.731&       -0.961\\
 $^{  147}_{   62}$&       2.311&    0&      18.528&      18.438& $^{  151}_{   64}$&       2.653&    0&      15.127&      15.621&$^{  151}_{   66}$&       4.180&    0&       4.283&       4.096\\
 $^{  153}_{   66}$&       3.559&    0&       8.389&       8.344&$^{  153}_{   68}$&       4.802&    0&       1.845&       1.756&$^{  155}_{   68}$&       4.118&    0&       6.160&       5.622\\
 $^{  155}_{   70}$&       5.338&    0&        0.304&        0.241&$^{  157}_{   70}$&       4.621&    0&       3.888&       3.753&$^{  157}_{   72}$&       5.880&    0&       -0.932&      -1.062\\
 $^{  159}_{   74}$&       6.450&    0&      -2.086&      -2.275&$^{  165}_{   74}$&       5.029&    2&       3.407&       5.090&$^{  161}_{   76}$&       7.066&    0&      -3.167&      -3.463\\
 $^{  167}_{   76}$&       5.980&    0&        0.225&        0.273& $^{  169}_{   76}$&       5.716&    0&       1.514&       1.347&$^{  173}_{   76}$&       5.055&    0&       5.028&       4.445\\
 $^{  167}_{   78}$&       7.160&    0&      -3.108&      -3.019&$^{  169}_{   78}$&       6.858&    0&      -2.156&      -2.077&$^{  171}_{   78}$&       6.610&    0&      -1.348&      -1.256\\
 $^{  173}_{   78}$&       6.350&    0&       -0.337&       -0.336&$^{  175}_{   78}$&       6.178&    2&       1.733&       1.739&$^{  177}_{   78}$&       5.643&    0&       2.326&       2.533\\
 $^{  181}_{   78}$&       5.150&    0&       4.865&       4.889&$^{  183}_{   78}$&       4.823&    0&       6.609&       6.672& $^{  173}_{   80}$&       7.378&    0&      -3.155&      -2.962\\
 $^{  175}_{   80}$&       7.043&    0&      -1.966&      -1.933&$^{  177}_{   80}$&       6.736&    2&       -0.928&        0.517&$^{  179}_{   80}$&       6.340&    0&        0.297&        0.519\\
 $^{  183}_{   80}$&       6.039&    0&       1.932&       1.679& $^{  185}_{   80}$&       5.774&    0&       2.928&       2.805&$^{  191}_{   82}$&       5.450&    0&       5.788&       5.267\\
 $^{  187}_{   84}$&       7.979&    2&       -0.854&      -1.930& $^{  189}_{   84}$&       7.701&    2&      -1.359&      -1.148&$^{  191}_{   84}$&       7.501&    0&       -0.831&      -1.987\\
 $^{  195}_{   84}$&       6.746&    0&        0.791&        0.530& $^{  197}_{   84}$&       6.412&    0&       2.086&       1.795&$^{  199}_{   84}$&       6.074&    0&       3.437&       3.189\\
 $^{  201}_{   84}$&       5.799&    0&       4.759&       4.415& $^{  205}_{   84}$&       5.324&    0&       7.195&       6.765&$^{  207}_{   84}$&       5.216&    0&       7.998&       7.332\\
 $^{  209}_{   84}$&       4.979&    2&      10.213&      10.118&$^{  211}_{   84}$&       7.595&    5&       -.283&       -0.849&$^{  213}_{   84}$&       8.536&    0&      -5.429&      -5.175\\
 $^{  195}_{   86}$&       7.690&    0&      -2.221&      -1.833& $^{  197}_{   86}$&       7.411&    0&      -1.268&       -0.974&$^{  199}_{   86}$&       7.140&    0&       -0.202&       -0.087\\
 $^{  201}_{   86}$&       6.861&    0&        0.959&        0.888&$^{  201}_{   86}$&       6.630&    0&       1.837&       1.769&$^{  207}_{   86}$&       6.251&    0&       3.422&       3.246\\
 $^{  209}_{   86}$&       6.156&    0&       4.010&       3.633& $^{  211}_{   86}$&       5.965&    2&       5.752&       5.901& $^{  213}_{   86}$&       8.244&    5&      -1.702&      -2.017\\
 $^{  215}_{   86}$&       8.839&    0&      -5.638&      -5.245& $^{  219}_{   86}$&       6.946&    2&        .698&       1.763&$^{  221}_{   86}$&       6.148&    2&       3.979&       4.939\\
 $^{  203}_{   88}$&       7.730&    0&      -1.509&      -1.276&$^{  205}_{   88}$&       7.490&    0&       -0.678&       -0.530&$^{  209}_{   88}$&       7.140&    0&        0.675&        0.618\\
 $^{  211}_{   88}$&       7.043&    0&       1.150&        0.941&$^{  211}_{   88}$&       6.861&    2&       2.658&       3.042& $^{  215}_{   88}$&       8.864&    5&      -2.810&      -2.969\\
 $^{  217}_{   88}$&       9.161&    0&      -5.796&      -5.353&$^{  219}_{   88}$&       8.138&    2&      -1.484&      -1.297&$^{  221}_{   88}$&       6.884&    2&       1.942&       2.823\\
 $^{  211}_{   90}$&       7.942&    0&      -1.432&      -1.259&$^{  213}_{   90}$&       7.837&    0&       -0.842&       -0.958& $^{  215}_{   90}$&       7.665&    2&        0.477&        0.983\\
 $^{  217}_{   90}$&       9.433&    5&      -3.577&      -3.697&$^{  219}_{   90}$&       9.510&    0&      -5.979&      -5.514& $^{  221}_{   90}$&       8.628&    2&      -2.353&      -1.949\\
 $^{  223}_{   90}$&       7.567&    2&       1.778&       1.202&$^{  225}_{   90}$&       6.920&    2&       3.810&       3.495&$^{  227}_{   90}$&       6.146&    2&       6.824&       6.747\\
 $^{  217}_{   94}$&       8.160&    0&      -1.796&       -0.467& $^{  219}_{   92}$&       9.940&    5&      -4.377&      -4.221&$^{  223}_{   92}$&       8.941&    0&      -4.260&      -3.495\\
 $^{  225}_{   92}$&       8.014&    0&      -1.080&       -.889&$^{  235}_{   94}$&       4.678&    1&      17.668&      17.560&$^{  239}_{   94}$&       5.245&    3&      15.404&      13.456\\
 $^{  241}_{   94}$&       5.140&    2&      15.712&      14.112&$^{  241}_{   96}$&       6.185&    3&      11.276&       9.199& $^{  243}_{   96}$&       6.169&    2&      10.784&       9.271\\
 $^{  245}_{   96}$&       5.622&    2&      13.661&      12.170&$^{  247}_{   96}$&       5.354&    1&      15.552&      13.903&$^{  249}_{   98}$&       6.296&    1&      11.653&       9.674\\
 $^{  251}_{   98}$&       6.177&    5&      12.037&      10.327&$^{  251}_{  100}$&       7.425&    1&       7.849&       5.617&$^{  253}_{  100}$&       7.199&    5&       8.210&       6.541\\
 $^{  255}_{  100}$&       7.240&    4&       8.014&       6.179&$^{  257}_{  100}$&       6.864&    2&       9.175&       7.685&$^{  263}_{  104}$&       8.250&    0&       3.301&       2.513\\
 $^{  259}_{  106}$&       9.804&    0&       -0.492&      -1.353& $^{  261}_{  106}$&       9.714&    0&       -0.638&      -1.140& $^{  269}_{  106}$&       8.700&    0&       2.079&       1.717\\
 $^{  271}_{  106}$&       8.670&    0&       2.219&       1.789&$^{  265}_{  108}$&      10.470&    0&      -2.699&      -2.436& $^{  267}_{  108}$&      10.037&    0&      -1.187&      -1.384\\
 $^{  273}_{  108}$&       9.730&    0&       -0.119&       -0.645&$^{  267}_{  110}$&      11.780&    0&      -5.553&      -4.723&$^{  269}_{  110}$&      11.509&    0&      -3.747&      -4.191\\
 $^{  271}_{  110}$&      10.899&    0&      -2.639&      -2.880&$^{  273}_{  110}$&      11.370&    0&      -3.770&      -3.945& $^{  277}_{  110}$&      10.720&    0&      -2.222&      -2.535\\
 $^{  281}_{  110}$&       9.320&    0&       2.125&       1.122& $^{  281}_{  112}$&      10.460&    0&      -1.000&      -1.309&$^{  285}_{  112}$&       9.320&    0&       1.447&       1.788\\
 $^{  287}_{  114}$&      10.170&    0&       -0.319&        0.032&$^{  289}_{  114}$&       9.980&    0&        0.279&        0.526& $^{  291}_{  116}$&      10.890&    0&      -1.721&      -1.206\\
 $^{  293}_{  116}$&      10.680&    0&      -1.276&       -0.708&\\
 

\hline
\end{tabular} }
\label{tab4}
\end{table*}

\begin{table*}
\hspace{0.2 cm}
	\caption{List of $\alpha$-decay half-lives for 84 o-e nuclei. Logarithm of predicted $\alpha$-decay half-lives $log_{10}(T_{1/2}^{pred.})$=$\tau^{pred.}$ in seconds using (eqn.32) with parameter fixed $c_f=0.19$, $r_0=0.98$ for $l=0$, $c_f=0.02$ for $l>0$. The experimental results of half-lives 
$log_{10}(T_{1/2}^{expt.})$=$\tau^{expt.}$ in seconds \cite{shan}. The $\alpha$-decay energies $Q_{\alpha}$ are in MeV.}
\renewcommand{\tabcolsep}{.1 cm}
\renewcommand{\arraystretch}{1.3}
{\begin{tabular}{|c|c|c|c|c|c|c|c|c|c|c|c|c|c|c|c|c|}
\hline
\cline{2-14}
\hline
Nucleus &       Q$_\alpha$ (MeV)  &  $l$&     $\tau^{expt.}$ (s) &  $\tau^{pred.}$ (s)&Nucleus &       Q$_\alpha$ (MeV)  &$l$&       $\tau^{expt.}$ (s) &  $\tau^{pred.}$ (s)&Nucleus &       Q$_\alpha$ (MeV) &$l$&        $\tau^{expt.}$ (s) &  $\tau^{pred.}$ (s)\\
\hline
 $^{  111}_{   53}$&       3.275&    0&       3.453&       2.321&$^{  147}_{   63}$&       2.991&    0&      10.976&      11.269& $^{  149}_{   65}$&       4.078&    2&       4.948&       5.632\\
 $^{  151}_{   65}$&       3.496&    2&       8.848&       9.684&$^{  153}_{   69}$&       5.248&    0&        0.211&        0.164&$^{  169}_{   77}$&       6.141&    0&        0.008&        0.069\\
 $^{  177}_{   77}$&       5.130&    0&       4.699&       4.524&$^{  173}_{   79}$&       6.836&    0&      -1.575&      -1.636& $^{  183}_{   79}$&       5.752&    0&       3.413&       2.446\\
 $^{  183}_{   79}$&       5.466&    0&       3.899&       3.779&$^{  185}_{   79}$&       5.180&    0&       4.992&       5.200&$^{  177}_{   81}$&       7.067&    0&      -1.608&      -1.623\\
 $^{  179}_{   81}$&       6.718&    0&       -0.638&       -0.449&$^{  187}_{   83}$&       7.779&    5&       -0.374&      -1.457&$^{  189}_{   83}$&       7.270&    5&       -0.046&        0.093\\
 $^{  193}_{   83}$&       6.304&    5&       4.849&       3.583&$^{  195}_{   83}$&       5.832&    5&       6.831&       5.620&$^{  211}_{   83}$&       6.750&    5&       2.187&       1.593\\
 $^{  213}_{   83}$&       5.988&    5&       5.145&       4.656& $^{  197}_{   83}$&       7.100&    0&       -0.411&      -1.175& $^{  199}_{   85}$&       6.777&    0&        0.886&        0.795\\
 $^{  201}_{   85}$&       6.473&    0&       2.148&       1.946&$^{  203}_{   85}$&       6.210&    0&       3.216&       3.012& $^{  205}_{   85}$&       6.020&    0&       4.208&       3.822\\
 $^{  207}_{   85}$&       5.872&    0&       4.879&       4.476&$^{  209}_{   85}$&       5.757&    0&       5.677&       4.998& $^{  211}_{   85}$&       5.982&    0&       4.793&       3.914\\
 $^{  213}_{   85}$&       9.254&    0&      -6.903&      -6.529& $^{  215}_{   85}$&       8.178&    0&      -4.000&      -3.895& $^{  217}_{   85}$&       7.201&    0&      -1.490&       -0.947\\
 $^{  219}_{   85}$&       6.324&    0&       1.761&       2.320&$^{  201}_{   87}$&       7.516&    0&      -1.161&       -0.966&$^{  203}_{   87}$&       7.260&    0&       -0.238&       -0.139\\
 $^{  205}_{   87}$&       7.055&    0&        0.593&        0.554& $^{  207}_{   87}$&       6.900&    0&       1.193&       1.094&$^{  209}_{   87}$&       6.777&    0&       1.754&       1.532\\
 $^{  211}_{   87}$&       6.663&    0&       2.330&       1.948&$^{  213}_{   87}$&       6.905&    0&       1.542&        0.996&$^{  215}_{   87}$&       9.540&    0&      -7.066&      -6.521\\
 $^{  217}_{   87}$&       8.469&    0&      -4.721&      -3.971&$^{  219}_{   87}$&       7.449&    0&      -1.694&       -0.982&$^{  221}_{   87}$&       6.458&    2&       2.547&       4.070\\
 $^{  223}_{   87}$&       5.562&    4&       7.530&       8.325&$^{  209}_{   89}$&       7.725&    0&      -1.009&       -0.945&$^{  211}_{   89}$&       7.620&    0&       -0.678&       -0.634\\
 $^{  213}_{   89}$&       7.501&    0&       -0.132&       -0.268& $^{  215}_{   89}$&       7.746&    0&       -0.767&      -1.090&$^{  217}_{   89}$&       9.832&    0&      -7.161&      -6.524\\
 $^{  219}_{   89}$&       8.830&    0&      -4.928&      -4.219&$^{  221}_{   89}$&       7.783&    4&      -1.117&        0.198& $^{  223}_{   89}$&       6.783&    2&       2.609&       3.620\\
 $^{  225}_{   89}$&       5.935&    2&       6.232&       7.300&$^{  227}_{   89}$&       5.042&    0&      11.019&      10.777&$^{  213}_{   91}$&       8.390&    0&      -2.276&      -2.231\\
 $^{  215}_{   91}$&       8.240&    0&      -1.854&      -1.824&$^{  217}_{   91}$&       8.489&    0&      -2.439&      -2.563&$^{  219}_{   91}$&      10.080&    0&      -7.276&      -6.432\\
 $^{  225}_{   91}$&       7.380&    2&        0.385&       2.222& $^{  227}_{   91}$&       6.580&    0&       3.731&       3.861& $^{  231}_{   91}$&       5.150&    0&      12.973&      11.133\\
 $^{  235}_{   93}$&       5.195&    1&      13.943&      13.455& $^{  237}_{   93}$&       4.958&    1&      15.452&      14.974& $^{  239}_{   95}$&       5.922&    1&      11.113&      10.259\\
 $^{  241}_{   95}$&       5.638&    1&      12.567&      11.779& $^{  243}_{   95}$&       5.439&    1&      13.986&      12.912&$^{  243}_{   97}$&       6.874&    2&       7.848&       6.497\\
 $^{  245}_{   97}$&       6.455&    2&       9.362&       8.335&$^{  249}_{   97}$&       5.525&    2&      13.610&      13.200&$^{  243}_{   99}$&       8.072&    0&       2.517&       1.416\\
 $^{  245}_{   99}$&       7.909&    3&       3.519&       3.320& $^{  255}_{  103}$&       8.556&    4&       3.095&       2.654&$^{  257}_{  103}$&       9.010&    0&        0.688&       -0.183\\
 $^{  257}_{  105}$&       9.206&    0&        0.389&       -0.033&$^{  259}_{  105}$&       9.620&    0&       -0.292&      -1.207&$^{  263}_{  105}$&       8.830&    0&       2.798&       1.012\\
 $^{  261}_{  107}$&      10.500&    0&      -1.899&      -2.781& $^{  267}_{  107}$&       9.230&    0&       1.230&        0.487&$^{  275}_{  109}$&      10.480&    0&      -1.699&      -2.258\\
 $^{  279}_{  111}$&      10.530&    0&      -1.046&      -1.782&$^{  283}_{  113}$&      10.480&    0&      -1.000&      -1.057& $^{  285}_{  113}$&      10.010&    0&        0.623&        0.152\\
 $^{  287}_{  115}$&      10.760&    0&      -1.432&      -1.158&$^{  289}_{  115}$&      10.520&    0&       -0.658&       -0.579&$^{  293}_{  117}$&      11.180&    0&      -1.854&      -1.611\\


\hline
\end{tabular}}
\label{tab5}
\end{table*}

\begin{table*}
\hspace{0.2 cm}
	\caption{List of $\alpha$-decay half-lives for 80 o-o nuclei. Logarithm of predicted $\alpha$-decay half-lives $log_{10}(T_{1/2}^{pred.})$=$\tau^{pred.}$  in seconds using (eqn.32) with parameter fixed $c_f=0.19$, $r_0=0.98$ for $l=0$, $c_f=0.02$ for $l>0$. The experimental results of half-lives 
$log_{10}(T_{1/2}^{expt.})$= $\tau^{expt.}$ in seconds \cite{shan}. The $\alpha$-decay energies $Q_{\alpha}$ are in MeV.}
\renewcommand{\tabcolsep}{.1 cm}
\renewcommand{\arraystretch}{1.3}
{\begin{tabular}{|c|c|c|c|c|c|c|c|c|c|c|c|c|c|c|c|c|}
\hline
\cline{2-14}
\hline
Nucleus &       Q$_\alpha$  (MeV)  &  $l$&     $\tau^{expt.}$ (s) &  $\tau^{pred.}$ (s)&Nucleus &       Q$_\alpha$(MeV)  &$l$&       $\tau^{expt.}$ (s) &  $\tau^{pred.}$ (s)&Nucleus &       Q$_\alpha$(MeV) &$l$&        $\tau^{expt.}$ (s) &  $\tau^{pred.}$ (s)\\
\hline
$^{  110}_{   53}$&       3.580&    0&        0.582&        0.349&$^{  112}_{   53}$&       2.990&    0&       5.455&       4.443&$^{  114}_{   55}$&       3.351&    0&       3.501&       3.098\\
 $^{  148}_{   63}$&       2.692&    0&      14.719&      14.407& $^{  152}_{   67}$&       4.507&    0&       3.130&       2.771& $^{  154}_{   67}$&       4.041&    0&       6.570&       5.529\\
 $^{  154}_{   69}$&       5.094&    0&       1.176&        0.852&$^{  156}_{   69}$&       4.345&    0&       5.117&       4.789&$^{  158}_{   71}$&       4.790&    0&       3.066&       3.373\\
 $^{  162}_{   73}$&       5.007&    5&       3.683&       5.130&$^{  160}_{   75}$&       6.697&    2&      -2.255&      -1.249&$^{  162}_{   75}$&       6.270&    0&       -0.957&      -1.231\\
 $^{  166}_{   77}$&       6.724&    0&      -1.947&      -2.011& $^{  170}_{   78}$&       7.177&    0&      -2.579&      -3.114&$^{  172}_{   79}$&       6.923&    0&      -1.658&      -1.914\\
 $^{  174}_{   79}$&       6.699&    0&       -0.811&      -1.178& $^{  182}_{   79}$&       5.526&    0&       4.759&       3.504&$^{  182}_{   81}$&       6.550&    0&       1.860&        0.124\\
 $^{  212}_{   83}$&       6.207&    5&       4.571&       3.720&$^{  214}_{   83}$&       5.621&    5&       7.163&       6.365&$^{  196}_{   85}$&       7.198&    0&       -0.385&       -0.662\\
 $^{  198}_{   85}$&       6.893&    0&        0.626&        0.381&$^{  200}_{   85}$&       6.596&    0&       1.878&       1.472&$^{  202}_{   85}$&       6.354&    0&       2.697&       2.418\\
 $^{  204}_{   85}$&       6.070&    0&       4.151&       3.611&$^{  208}_{   85}$&       5.751&    0&       6.044&       5.041&$^{  210}_{   85}$&       5.631&    2&       7.741&       7.040\\
 $^{  212}_{   85}$&       7.817&    5&       -0.423&      -1.150&$^{  214}_{   85}$&       8.987&    0&      -6.249&      -5.932&$^{  216}_{   85}$&       7.950&    0&      -3.512&      -3.268\\
 $^{  218}_{   85}$&       6.874&    0&       1.620&        0.188& $^{  200}_{   87}$&       7.621&    0&      -1.310&      -1.289&$^{  202}_{   87}$&       7.389&    0&       -0.429&       -0.562\\
 $^{  204}_{   87}$&       7.171&    0&        0.434&        0.156&$^{  206}_{   87}$&       6.923&    0&       1.280&       1.022&$^{  208}_{   87}$&       6.772&    0&       1.822&       1.564\\
 $^{  210}_{   87}$&       6.672&    2&       2.429&       3.351&$^{  212}_{   87}$&       6.529&    2&       4.106&       3.897&$^{  214}_{   87}$&       8.589&    5&      -2.270&      -2.599\\
 $^{  216}_{   87}$&       9.174&    0&      -6.133&      -5.711&$^{  218}_{   87}$&       8.014&    0&      -2.964&      -2.717&$^{  220}_{   87}$&       6.801&    1&       1.620&       2.876\\
 $^{  206}_{   89}$&       7.945&    0&      -1.658&      -1.589& $^{  208}_{   89}$&       7.730&    0&      -1.018&       -0.948& $^{  210}_{   89}$&       7.607&    0&       -0.456&       -0.578\\
 $^{  212}_{   89}$&       7.159&    0&       -0.018&        0.930& $^{  214}_{   89}$&       7.352&    2&       1.232&       1.643&$^{  216}_{   89}$&       9.235&    5&      -3.311&      -3.551\\
 $^{  218}_{   89}$&       9.380&    0&      -5.967&      -5.537&$^{  222}_{   89}$&       7.137&    0&        0.730&        0.879&$^{  224}_{   89}$&       6.327&    1&       5.722&       5.657\\
 $^{  212}_{   91}$&       8.429&    0&      -2.292&      -2.329& $^{  216}_{   91}$&       8.097&    0&       -0.449&      -1.411& $^{  226}_{   91}$&       6.987&    0&       2.454&       2.236\\
 $^{  230}_{   91}$&       5.439&    2&      11.308&      10.850&$^{  254}_{   99}$&       6.616&    1&      10.424&       8.552&$^{  256}_{  101}$&       7.856&    4&       6.710&       4.234\\
 $^{  256}_{  105}$&       9.340&    0&        0.359&       -0.402&$^{  258}_{  105}$&       9.500&    0&        0.776&       -0.870&$^{  260}_{  107}$&      10.400&    0&      -1.456&      -2.530\\
 $^{  264}_{  107}$&       9.960&    0&       -0.357&      -1.483&$^{  266}_{  107}$&       9.430&    0&        0.230&       -0.076&$^{  270}_{  107}$&       9.060&    0&       1.785&        0.956\\
 $^{  272}_{  107}$&       9.310&    0&       1.000&        0.194&$^{  274}_{  107}$&       8.930&    0&       1.732&       1.304&$^{  268}_{  109}$&      10.670&    0&      -1.678&      -2.628\\
 $^{  274}_{  109}$&      10.200&    0&       -0.357&      -1.552& $^{  276}_{  109}$&      10.030&    0&       -0.347&      -1.139&$^{  278}_{  109}$&       9.580&    0&        0.653&        0.054\\
 $^{  272}_{  111}$&      11.197&    0&      -2.420&      -3.254&$^{  274}_{  111}$&      11.480&    0&      -2.194&      -3.890&$^{  278}_{  111}$&      10.850&    0&      -2.377&      -2.536\\
 $^{  280}_{  111}$&       9.910&    0&        0.663&       -0.194& $^{  278}_{  113}$&      11.850&    0&      -3.620&      -4.118& $^{  182}_{  113}$&      10.780&    0&      -1.155&       -0.400\\
 $^{  284}_{  113}$&      10.120&    0&       -0.041&       -0.133& $^{  286}_{  113}$&       9.790&    0&        0.978&        0.750&$^{  288}_{  115}$&      10.630&    0&      -1.060&       -0.846\\
 $^{  290}_{  115}$&      10.410&    0&       -0.187&       -0.307& $^{  294}_{  117}$&      11.070&    0&      -1.097&      -1.361\\
 

\hline
\end{tabular} }
\label{tab6}
\end{table*}

\begin{table}
\hspace{0.2 cm}
	\caption{The $Q_{\alpha}$ represent the $\alpha$-decay energies  in MeV and the experimental results of half-lives $log_{10}(T_{1/2}^{expt.})$=$\tau^{expt.}$ in seconds \cite{mohr}. Logarithm of predicted $\alpha$-decay half-lives 
$log_{10}(T_{1/2}^{pred.})$=$\tau^{pred.}$ in seconds using (eqn.32) with parameter fixed $c_f=0.19$, $r_0=0.98$ for $l=0$ of chain1 and chain2 }
\renewcommand{\tabcolsep}{.2cm}
\renewcommand{\arraystretch}{1.1}
{\begin{tabular}{|c c c c c|}
\hline
\cline{2-5}
\hline
 &Nucleus &       Q$_\alpha$ (MeV)  &     $\tau^{expt.}$ (s) &  $\tau^{pred.}$ (s) \\
\hline
chain1&$^{286}Fl\rightarrow^{282}Cn$ &       10.35 &    -0.699  &       -0.432\\
chain1&$^{290}Lv\rightarrow^{286}Fl$ &       11.0  &    -2.081  &       -1.460\\
chain1&$^{294}118\rightarrow^{290}Lv$&       11.82 &    -3.161  &       -2.767\\
chain2&$^{288}Fl\rightarrow^{284}Cn$ &       10.07 &    -0.180  &        0.290\\
chain2&$^{292}Lv\rightarrow^{288}Fl$ &       10.78 &    -1.886  &       -0.946\\
chain2&$^{296}118\rightarrow^{292}Lv$&       11.655&    -3.083 &        -2.428\\


\hline
\end{tabular} }
\label{tab7}
\end{table}

\begin{figure}[ht]
\includegraphics[width=1.2\columnwidth]{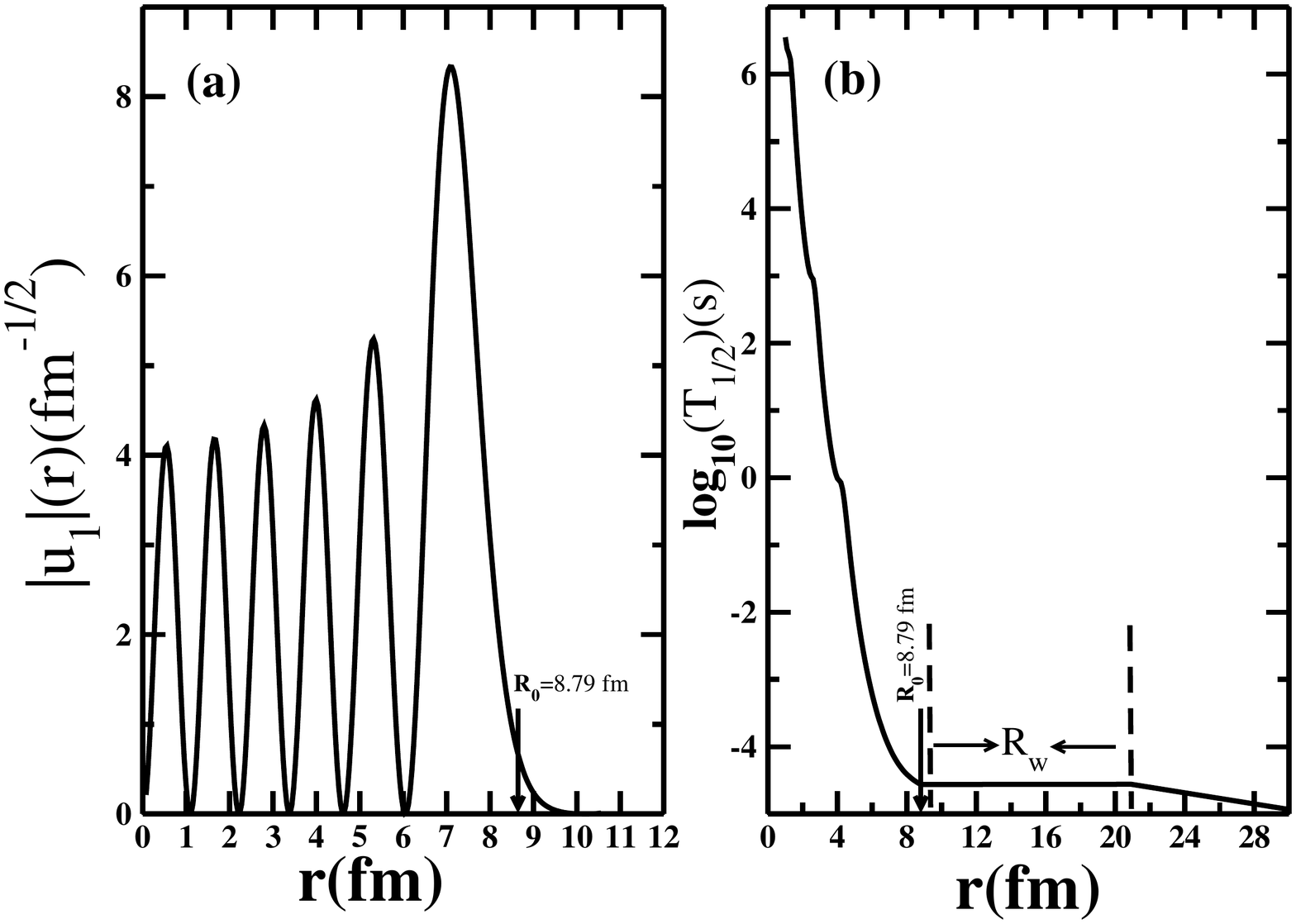}
	\caption{Plots of (a) the modulous of the radial wave function $|u_1(r)|$ at resonance as a function of r, (b) alpha decay half-lives 
$log_{10}(T_{1/2}^{calt.}) $ (eqn.7) as a function of $r$ for $^{106}Te$ with $R_0$=8.79 fm and $R_w$ showing the radial independence region having range 8.9-21.0 fm. }
\label{fig2}
\end{figure}

\begin{figure}[ht]
\includegraphics[width=1.2\columnwidth]{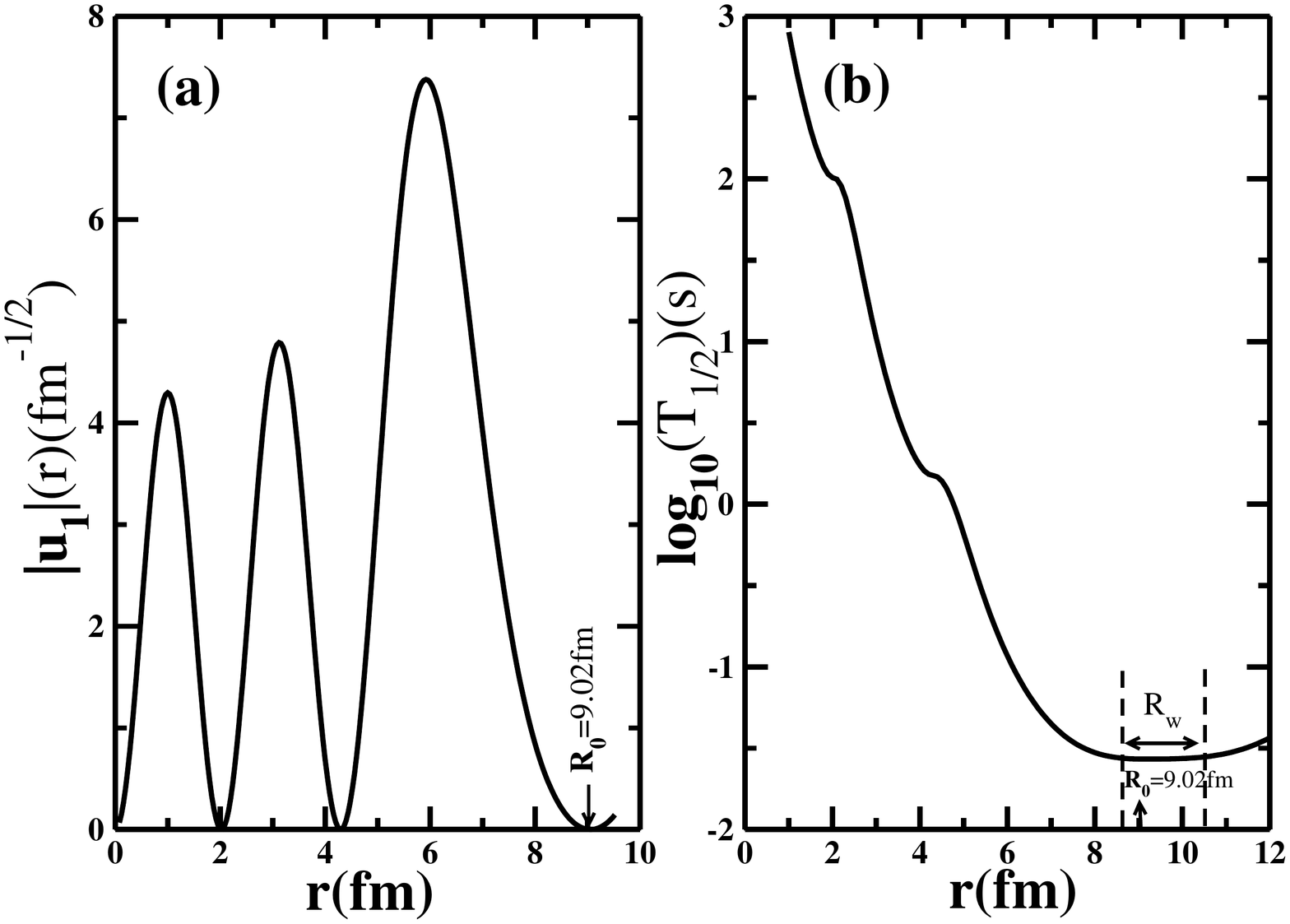}
	\caption{Plots of (a) the modulous of the radial wave function $|u_1(r)|$ at resonance as a function of r, (b) proton decay half-lives $log_{10}(T_{1/2}^{calt.}) $ (eqn.7) as a function of $r$ for $^{167}$Ir with $R_0$=9.02 fm and $R_w$ showing the radial independence region having range 8.6-10.5 fm.}
\label{fig3}
\end{figure}

\begin{figure}[ht]
\includegraphics[width=1.0\columnwidth]{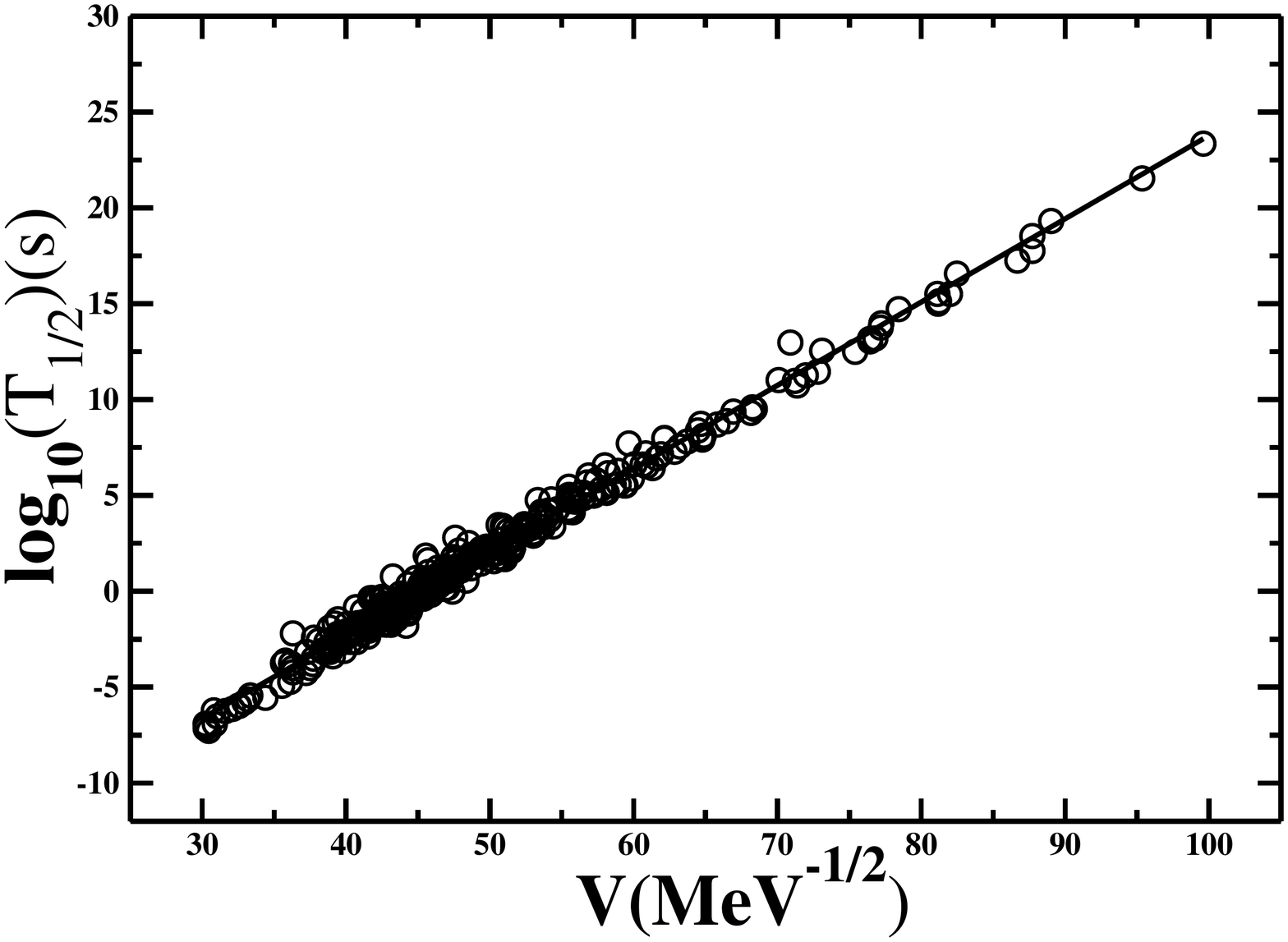}
	\caption{Plot of $\alpha$-decay half-lives $log_{10} (T_{1/2}^{expt.})$ from experiments (solid dots) and $log_{10} (T_{1/2}^{pred.})$ (eqn.32) as a function of $V=a\chi + c$ for $l=0$ state with Z=52-118. }
\label{fig4}
\end{figure}

\begin{figure}[ht]
\includegraphics[width=1.0\columnwidth]{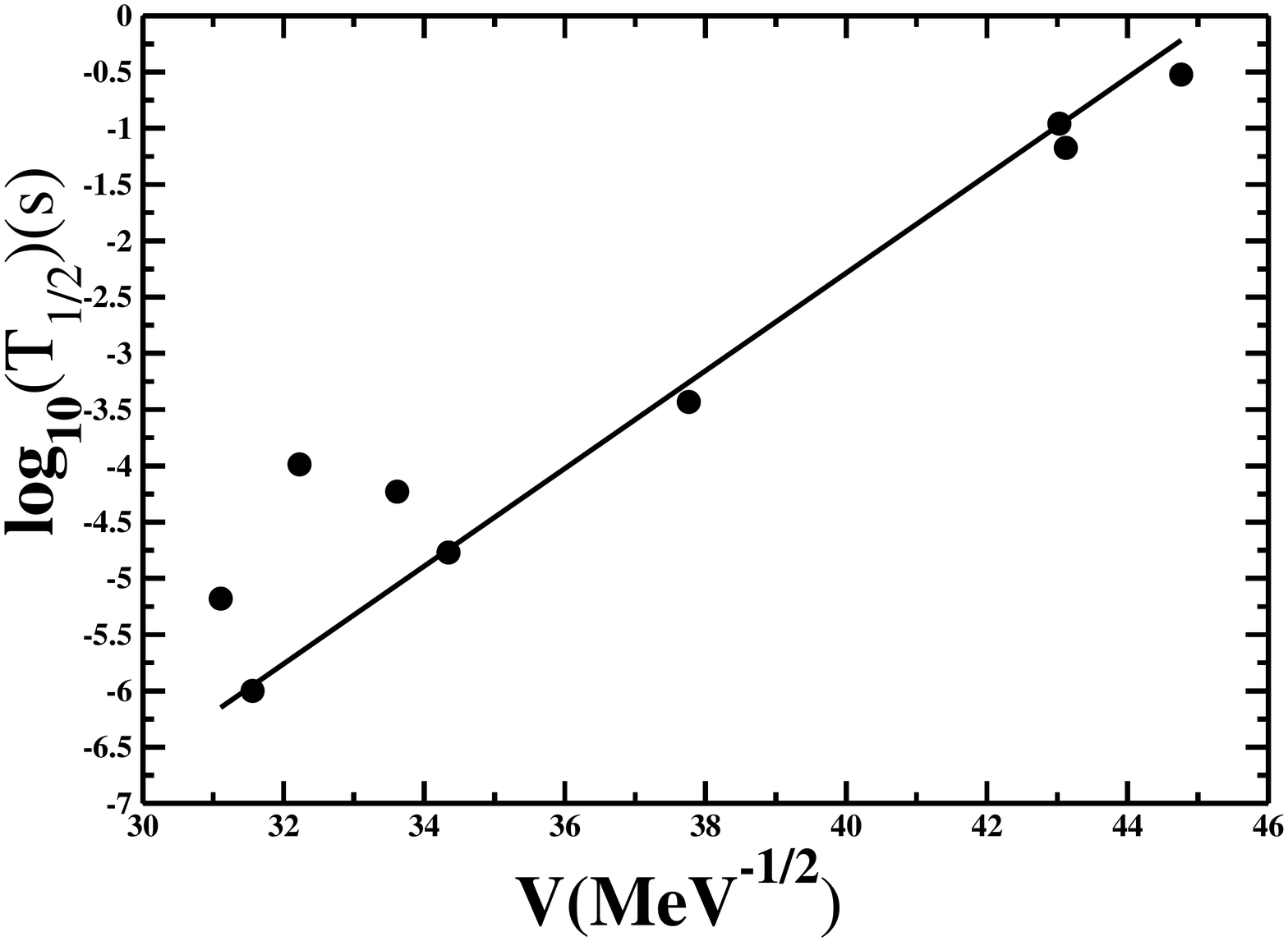}
\caption{Plot of proton-decay half-lives $log_{10} (T_{1/2}^{expt.})$ from experiments (solid dots) and $log_{10} (T_{1/2}^{pred.})$ (eqn.32) as a function of $V=a\chi + c$ for $l=0$ state. }
\label{fig5}
\end{figure}

\begin{table*}
\hspace{0.2 cm}
\caption{The $\alpha$-decay energies $Q^{FRDM}$ in MeV and the half-lives 	$log_{10}(T_{1/2}^{FRDM})$=$\tau^{FRDM}$ in seconds  for Z=119 and Z=120 systems are taken from Finite range droplet model (FRDM) \cite{frdm}. Logarithm of predicted $\alpha$-decay half-lives $log_{10}(T_{1/2}^{pred.})$=$\tau^{pred.}$ in seconds using (eqn.32) with parameter fixed $c_f=0.19$, $r_0=0.98$ for $l=0$.}
\renewcommand{\tabcolsep}{.05cm}
\renewcommand{\arraystretch}{1.4}
{\begin{tabular}{|c|c|c|c|c|c|c|c|c|c|c|c|c|c|}
\hline
\cline{2-11}
\hline
Nucleus &       Q$^{FRDM}$  (MeV)  &     $\tau^{FRDM}$ (s) &  $\tau^{pred.}$ (s)&Nucleus &       Q$^{FRDM}$(MeV)  &       $\tau^{FRDM}$ (s) &  $\tau^{pred.}$ (s)&Nucleus &       Q$^{FRDM}$(MeV) &        $\tau^{FRDM}$ (s) &  $\tau^{pred.}$ (s)\\

\hline
 $^{ 284}_{ 119}$&      13.020&      -4.420&      -4.775& $^{ 285}_{ 119}$&      13.790&      -6.250&      -6.148 & $^{ 286}_{ 119}$&      13.740&      -5.810&      -6.075\\
 $^{ 287}_{ 119}$&      13.450&      -5.600&      -5.587&$^{ 288}_{ 119}$&      13.380&      -5.140&      -5.475&$^{ 289}_{ 119}$&      13.550&      -5.790&      -5.784\\
 $^{ 290}_{ 119}$&      13.360&      -5.090&      -5.462&$^{ 291}_{ 119}$&      13.200&      -5.130&      -5.186& $^{ 292}_{ 119}$&      13.170&      -4.710&      -5.143\\
$^{ 293}_{ 119}$&      12.880&      -4.470&      -4.617&$^{ 294}_{ 119}$&      12.800&      -3.970&      -4.476&$^{ 295}_{ 119}$&      12.880&      -4.490&      -4.640\\
$^{ 296}_{ 119}$&      13.080&      -4.540&      -5.024& $^{ 297}_{ 119}$&      12.740&      -4.190&      -4.395& $^{ 298}_{ 119}$&      12.500&      -3.340&      -3.936\\
$^{ 299}_{ 119}$&      12.800&      -4.320&      -4.533&$^{ 300}_{ 119}$&      13.150&      -4.670&      -5.197&$^{ 301}_{ 119}$&      13.270&      -5.250&      -5.425\\
$^{ 302}_{ 119}$&      13.380&      -5.130&      -5.632&$^{ 303}_{ 119}$&      13.380&      -5.460&      -5.643&$^{ 304}_{ 119}$&      14.140&      -6.550&      -6.932\\
$^{ 305}_{ 119}$&      13.840&      -6.330&      -6.453&$^{ 306}_{ 119}$&      13.970&      -6.230&      -6.678&$^{ 307}_{ 119}$&      13.810&      -6.290&      -6.425\\
$^{ 308}_{ 119}$&      13.430&      -5.230&      -5.785&$^{ 309}_{ 119}$&      13.310&      -3.350&      -5.584&$^{ 310}_{ 119}$&      12.760&      -3.880&      -4.578\\
$^{ 311}_{ 119}$&      12.490&      -3.650&      -4.062&$^{ 312}_{ 119}$&      12.110&      -2.480&      -3.296&$^{ 313}_{ 119}$&      13.250&      -5.220&      -5.520\\
$^{ 314}_{ 119}$&       4.660&      20.000&      29.878&$^{ 315}_{ 119}$&       4.130&      20.000&      35.493&$^{ 316}_{ 119}$&       5.140&      20.000&      25.546\\
$^{ 317}_{ 119}$&       8.140&       9.110&       8.204&$^{ 318}_{ 119}$&       7.840&      10.740&       9.442&$^{ 319}_{ 119}$&       8.740&       6.790&       5.891\\
$^{ 320}_{ 119}$&       8.680&       7.360&       6.097&$^{ 321}_{ 119}$&       8.710&       6.920&       5.976&$^{ 322}_{ 119}$&       8.720&       7.210&       5.929\\
$^{ 323}_{ 119}$&       8.980&       5.940&       4.999&$^{ 324}_{ 119}$&       9.000&       6.210&       4.919& $^{ 325}_{ 119}$&       8.940&       6.090&       5.115\\
$^{ 326}_{ 119}$&       8.720&       7.210&       5.883&$^{ 327}_{ 119}$&       8.550&       7.520&       6.496&$^{ 328}_{ 119}$&       8.270&       8.950&       7.557\\
$^{ 329}_{ 119}$&       8.080&       9.350&       8.308&$^{ 330}_{ 119}$&       6.240&      18.920&      17.513&$^{ 331}_{ 119}$&       6.030&      19.870&      18.828\\
$^{ 332}_{ 119}$&       5.980&      20.000&      19.143&$^{ 333}_{ 119}$&       7.820&      10.480&       9.353&$^{ 334}_{ 119}$&       7.570&      11.900&      10.446\\
$^{ 335}_{ 119}$&       7.420&      12.280&      11.126&$^{ 336}_{ 119}$&       7.160&      13.860&      12.366& $^{ 337}_{ 119}$&       6.760&      15.580&      14.428\\
$^{ 338}_{ 119}$&       6.540&      17.110&      15.641&$^{ 339}_{ 119}$&       6.360&      17.820&      16.681&$^{ 287}_{ 120}$&      13.980&      -6.070&      -6.230\\
$^{ 288}_{ 120}$&      13.920&      -7.020&      -6.142&$^{ 289}_{ 120}$&      13.890&      -5.890&      -6.103&$^{ 290}_{ 120}$&      13.770&      -6.750&      -5.912\\
$^{ 291}_{ 120}$&      13.910&      -5.930&      -6.159&$^{ 292}_{ 120}$&      13.890&      -6.960&      -6.137&$^{ 293}_{ 120}$&      13.690&      -5.510&      -5.809\\
$^{ 294}_{ 120}$&      13.290&      -5.820&      -5.115&$^{ 295}_{ 120}$&      13.360&      -4.890&      -5.252&$^{ 296}_{ 120}$&      13.690&      -6.590&      -5.842\\
$^{ 297}_{ 120}$&      13.540&      -5.230&      -5.594&$^{ 298}_{ 120}$&      13.350&      -5.930&      -5.268&$^{ 299}_{ 120}$&      13.100&      -4.370&      -4.823\\
$^{ 300}_{ 120}$&      13.400&      -6.030&      -5.380&$^{ 301}_{ 120}$&      13.670&      -5.480&      -5.864&$^{ 302}_{ 120}$&      13.720&      -6.630&      -5.960\\
$^{ 303}_{ 120}$&      13.840&      -5.800&      -6.175&$^{ 304}_{ 120}$&      13.820&      -6.830&      -6.152&$^{ 305}_{ 120}$&      14.480&      -6.940&      -7.230\\
$^{ 306}_{ 120}$&      14.270&      -7.640&      -6.911&$^{ 307}_{ 120}$&      14.140&      -6.350&      -6.713&$^{ 308}_{ 120}$&      14.300&      -7.700&      -6.980\\
$^{ 309}_{ 120}$&      13.730&      -5.590&      -6.054&$^{ 310}_{ 120}$&      13.630&      -6.480&      -5.893&$^{ 311}_{ 120}$&      13.110&      -4.390&      -4.975\\
$^{ 312}_{ 120}$&      12.670&      -4.550&      -4.148&$^{ 313}_{ 120}$&      12.260&      -2.580&      -3.332&$^{ 314}_{ 120}$&      11.590&      -2.120&      -1.884\\
$^{ 315}_{ 120}$&       4.660&      20.000&      30.465&$^{ 316}_{ 120}$&       4.160&      20.000&      35.779\\
\hline
\end{tabular}}
\label{tab8}
\end{table*}
\begin{table*}
\hspace{0.2 cm}
	\caption{The proton-decay energies $Q_{p}$ in MeV  and the experimental results of half-lives $log_{10}(T_{1/2}^{expt.})$=$\tau^{expt.}$ in seconds \cite{bsahua}. Logarithm of predicted proton-decay half-lives $log_{10}(T_{1/2}^{pred.})$=$\tau^{pred.}$ in seconds using (eqn.32) with parameter fixed $c_f=2.8$, $r_0=0.97$ for $l=0$ and $c_f=0.45$, $r_0=0.97$ for $l>0$.  }
\renewcommand{\tabcolsep}{.1cm}
\renewcommand{\arraystretch}{1.3}
{\begin{tabular}{|c|c|c|c|c|c|c|c|c|c|c|c|c|c|c|c|c|}
\hline
\cline{2-14}
\hline
Nucleus &       Q$_p$  (MeV)  & $l$&    $\tau^{expt.}$ (s) &  $\tau^{pred.}$ (s)&Nucleus &       Q$_p$(MeV)  &  $l$ &    $\tau^{expt.}$ (s) &  $\tau^{pred.}$ (s)&Nucleus &       Q$_p$(MeV) &  $l$&      $\tau^{expt.}$ (s) &  $\tau^{pred.}$ (s)\\

\hline
  $^{ 105}$Sb&        0.491&    2&       2.049&       2.300&
 $^{ 109}$I&        0.819&    0&      -3.987&      -5.661&
 $^{ 112}$Cs&        0.814&    2&      -3.301&      -2.985\\

 $^{ 113}$Cs&        0.973&    2&      -4.777&      -5.159&
 $^{ 117}$La&        0.803&    2&      -1.628&      -2.045&
 $^{ 117}$La&        0.954&    5&      -2.000&      -1.395\\

 $^{ 131}$Eu&        0.940&    2&      -1.749&      -1.975&
 $^{ 140}$Ho&       1.094&    3&      -2.221&      -2.098&
 $^{ 141}$Ho&       1.177&    3&      -2.387&      -3.072\\

 $^{ 141}$Ho&       1.256&    0&      -5.180&      -6.148&
 $^{ 145}$Tm&       1.753&    5&      -5.409&      -5.472&
 $^{ 146}$Tm&       1.127&    5&      -1.096&        0.018\\

 $^{ 146}$Tm&       1.307&    5&       -0.698&      -1.960&
 $^{ 147}$Tm&       1.071&    5&        0.591&        0.723&
 $^{ 147}$Tm&       1.139&    2&      -3.444&      -2.610\\

 $^{ 150}$Lu&       1.283&    5&      -1.180&      -1.179&
 $^{ 150}$Lu&       1.317&    2&      -4.523&      -3.923&
 $^{ 151}$Lu&       1.255&    5&       -0.896&       -0.892\\

 $^{ 151}$Lu&       1.332&    2&      -4.796&      -4.075&
 $^{ 155}$Ta&       1.791&    5&      -4.921&      -4.866&
 $^{ 156}$Ta&       1.028&    2&       -.620&        0.158\\

 $^{ 156}$Ta&       1.130&    5&        0.949&       1.146&
 $^{ 157}$Ta&        0.947&    0&       -0.523&       -0.218&
 $^{ 160}$Re&       1.284&    2&      -3.046&      -2.430\\

 $^{ 161}$Re&       1.338&    5&       -0.488&       -0.697&
 $^{ 161}$Re&       1.214&    0&      -3.432&      -3.258&
 $^{ 164}$Ir&       1.844&    5&      -3.959&      -4.356\\

 $^{ 165}$Ir&       1.546&    0&      -6.000&      -5.951&
 $^{ 165}$Ir&       1.733&    5&      -3.469&      -3.602&
 $^{ 166}$Ir&       1.168&    2&       -0.824&       -0.470\\

 $^{ 166}$Ir&       1.340&    5&       -0.076&       -0.191&
 $^{ 167}$Ir&       1.086&    0&       -0.959&       -0.970&
 $^{ 167}$Ir&       1.261&    5&        0.875&        0.674\\

 $^{ 171}$Au&       1.469&    0&      -4.770&      -4.741&
 $^{ 171}$Au&       1.718&    5&      -2.654&      -3.055&
 $^{ 177}$Tl&       1.180&    0&      -1.174&       -0.930\\

 $^{ 177}$Tl&       1.986&    5&      -3.347&      -4.454&
 $^{ 185}$Bi&       1.624&    0&      -4.229&      -5.058\\

\hline
\end{tabular}}
\label{tab9}
\end{table*}
\section{Results and Discussions:} \label{sec3}
In our present calculation of $\alpha$-decay half-lives, we take the exactly solvable potential as described in (eqn.13) and denote it as the effective Coulomb-nuclear potential $V_{eff}$ for the $\alpha+$nucleus system and change only the steepness of the interior side of the barrier i.e. $d_1$ but keep $d_2$ fixed. With the Coulomb function $F_l(r)$ and the wave function $u_1(r)$ at resonance, we calculate the half-life $T_{1/2}^{calt.}$ by using (eqn.7).

Now we plot $log_{10}(T_{1/2}^{calt.})$ as a function of r. Astonishingly we get to know that the $log_{10}(T_{1/2}^{calt.})$ values vary from infinity to lesser values with increase in r, become constant at a particular region and finally its values decreases for larger radial distance. To sum up, we can say that for a particular system there is a region where $log_{10}(T_{1/2}^{calt.})$ is constant and thus in that region radial dependence is removed. Moreover, for each system the constant stability radial region is changing a little bit but more or less it is coming out to be in the range 8.9 fm to 21 fm. Fig. \ref{fig2} shows the variation of $log_{10}(T_{1/2}^{calt.})$ vs r for the $^{106}Te$ nucleus where the radial independence region is coming out to be $R_w$=8.9-21.0 fm for $\alpha$-decay. In the same way, for proton decay we plot the $log_{10}(T_{1/2}^{calt.})$ vs r for the $^{  167}$Ir nucleus where the radial independence region is coming out to be $R_w$=8.6-10.5 fm as shown in Fig. \ref{fig3}. A close observation on $R_w$ both for $\alpha$-decay as well as proton-decay suggest that the constancy region is more prominently wider in case of $\alpha$-decay with $R_w$=8.9-21.0 fm than proton decay with $R_w$=8.6-10.5 fm. The radial distance where the constancy creeps on is the region where the wavefunction $u_1(r)$ vanishes to zero. In other words, the wave function $u_1(r)$ decreases rapidly with distance outside the nucleus. The confinement of the wave at resonance energy is the significance of molecular state with large amplitude inside and small or negligible amplitude outside the barrier. Thus the radial dependence restriction is removed and we have a solid proof that for a particular system there is a region where $log(T_{1/2})$ is remaining constant.

Since radial constancy is proved we henceforth take the $R_0$ value for calculation of half-lives for a chain of nuclei. To do that, we condense the integral J given by (eqn.8) and write in terms of $c_f$, $x_m$ and $F_l^{ps}$ as mentioned in (eqn.30). Our analysis show that for different nuclei the values of $c_f$ comes out to be in the range 0.139 to 0.37  for $\alpha$-decay and in the range 2.52 to 4.56 for proton-decay for $l=0$. For $\alpha$-decay half-life calculation, we take $c_f$=0.19 for $l=0$ and $c_f$=0.02 for $l>0$ and for proton-decay half-life calculation, we take $c_f$=2.8 for $l=0$ and $c_f$=0.45 for $l>0$ to maintain uniformity in all sets of nuclei. Furthermore, using this $c_f$, we estimate the values of $T_{1/2}$ by using the closed form expression (eqn.32) for the decimal logarithm of half-life and represent it as $log_{10} (T_{1/2}^{(pred.)})
=log (T_{1/2}^{pred.})/2.30258$. The $Q$ values, the  corresponding $d_1$ determining the steepness of the inner side of the barrier, the experimental half-lives $log_{10} (T_{1/2}^{expt.})$, the calculated half-lives $log_{10} (T_{1/2}^{calt.})$, the predicted half-lives $log_{10} (T_{1/2}^{pred.})$, the barrier radius $R_0$ and the $R_w$ values have been shown for a series of systems in Table \ref{tab1} for $\alpha$-decay and in Table \ref{tab2} for proton decay.
Analyzing the constancy region $R_w$ for all the systems mentioned in Table 
\ref{tab1}, it is seen that for the case of $\alpha$-decay the radial independence starts from ($R_0$+0.04) to near about 20 fm when $r_0$=0.97. To overcome this problem, we take the $r_0$ value to be 0.98 instead of 0.97 so as to make $R_0$ fall in the constancy region $R_w$ for all the systems and then predict the half-lives. But for the case of proton-decay we go with $r_0$=0.97 as in all the systems considered in Table \ref{tab2}, $R_0$ values come in the range of $R_w$.

 We then compare the experimental $\alpha$-decay results $log_{10}(T_{1/2}^{expt.})$ and predicted $log_{10}(T_{1/2})$ values i.e $log_{10}(T_{1/2}^{pred.})$ 
 (eqn.32) and present systematically in Table \ref{tab3}, \ref{tab4}, \ref{tab5} and \ref{tab6} for 144 e-e nuclei, 112 e-o nuclei, 84 o-e nuclei and 80 o-o nuclei respectively by taking $c_f$=0.19 for $l=0$ and $c_f$=0.02 for $l>0$ cases. For a large assemblage of nuclei starting from Z=52 to 118, we write the $log_{10}(T_{1/2})$ values viz. $log_{10}(T_{1/2}^{expt.})$, $log_{10}(T_{1/2}^{pred.})$.

 As we know the quest for synthesizing superheavy elements in the laboratory is still on. Nuclei up to Z=118 has been synthesized already \cite{yu1,yu2} 
and consistent endeavours have been there for synthesizing Z=119,120 nuclei. Thus, there is an acute necessity for the theoretical predictions of half-lives and other properties of these superheavy nuclei. Table \ref{tab7} presents the predicted results of $\alpha$-decay half-lives for two chains, chain1 and chain2 containing the Z=118 nucleus and the predicted half-lives are coming close to the experimental ones. Specifically in chain1 for the decay of $^{294}118\rightarrow\;^{290}Lv$ the $log_{10}(T_{1/2}^{expt.})$= -3.161 s and $log_{10}(T_{1/2}^{pred.})$= -2.767 s. Similarly in chain2 for the decay of $^{296}118\rightarrow\;^{292}Lv$ the $log_{10}(T_{1/2}^{expt.})$= -3.083 s and $log_{10}(T_{1/2}^{pred.})$= -2.428 s.
The predicted $\alpha$-decay half-lives for a series of nuclei with Z=119 and 
Z=120 are compared with the corresponding half-lives obtained from Finite range droplet model (FRDM) and shown in Table \ref{tab8}. In the table the 
$log_{10}(T_{1/2}^{expt.})$ and $log_{10}(T_{1/2}^{pred.})$ values are listed for $284 \le A \le 339$  for Z=119 nuclei and $287 \le A \le 316$ for Z=120 nuclei yielding 
satisfactory results. 

 Also a separate list of proton-decay half-lives are shown in Table \ref{tab9} by taking $c_f$=2.8 for $l=0$ and $c_f$=0.45 for $l>0$. Our findings reveal that we get a wide band of $log_{10}(T_{1/2}^{pred.})$ ranging from decimal logarithmic values of -6.529 s to 23.604 s. By considering all the systems listed in 
 Table \ref{tab3}, \ref{tab4}, \ref{tab5} and \ref{tab6} for the $\alpha$-decay half-lives the average deviation and standard deviation are found to be 0.481 and 0.653 respectively. Similarly, for the proton-decay half-lives listed in 
 Table \ref{tab9}, the average deviation and standard deviation come out to be 0.436 and 0.590 respectively. 

Now to clarify the ambiguity on the non linearity of Geiger-Nuttall law for various $\alpha$-emitters, we make a plot of $log_{10}(T_{1/2}^{expt.})$ and $log_{10}(T_{1/2}^{pred.})$ as a function of $V=a\chi+c$ used in (eqn.32) for $l=0$ and present in Fig. \ref{fig4} for $\alpha$-decay and in Fig. \ref{fig5} for proton-decay. To our utter surprise, we find that the plots of  $log_{10}(T_{1/2}^{expt.})$ and $log_{10}(T_{1/2}^{pred.})$  vs $V=a\chi+c$ give a single straight line. This clearly indicate that the our measured results have great accuracy. Finally we close this section with the physical
significance of $R_0$. 

{\it The physical significance of $R_0$ are as follows:}

1. The $R_0$ value we are considering here is coming close to the Global formula; $R_B=r_B(A_1^{1/3}+A_2^{1/3})+2.72$ \cite{brog}, $r_B=1.07$ where $R_B$ is the potential barrier radius. To illustrate this, we consider one of the systems of Table \ref{tab1} i.e. $\alpha$-decay of $^{ 148}_{  64}$ system. Here $R_0$=9.34 fm, $R_w$ is 9.2-15.0 fm and $R_B$=10.02 fm. Similarly in Table \ref{tab2} for proton decay, for the case of $^{  167}$Ir system $R_0$=9.02 fm, $R_w$ is 8.6-10.5 fm and $R_B$=9.67 fm. The mentioned values indicate the closeness of $R_0$ and $R_B$.

2. The wave function at resonance is negligibly small at $r\approx R_0$ which is very close to $R_B$. For example, in Fig. \ref{fig2} i.e for $^{ 106}Te$ system the wave function diminishes at $R_0$=8.79 fm and in Fig. \ref{fig3} i.e. for $^{  167}$Ir system the wave function diminishes at $R_0$=9.02 fm. Thus, we can say that the wave function is dying under the barrier.
 
3. In the radial variation of time, there is a region $R_w$ where $log_{10}(T_{1/2}^{calt.})$ is remaining constant and the $R_0$ which we have used in our calculation is found in that region.

\section{Summary and Conclusions}\label{conc}

By using the regular Coulomb function, resonant wave function and the difference in potentials a general formula is being put forth for the calculation of $\alpha$-decay width. The $\alpha$+nucleus potential is represented by special expressions of the potential. We have also seen that from the logarithm of half-lives vs radial distance plot, a radial independence region can be traced where the half-life is remaining constant. Thus for each of the systems a radial independence region can be tracked down and a particular radial distance falling in this independence zone can be used in the closed form expression. This derived formula is impeccable in predicting the $\alpha$-decay and proton-decay half-lives of any nuclei. Specifically for Z=118 nuclei on which many works have been carried out recently and the predicted half-lives are coming close to the experimental ones. Isotopes of nuclei with Z=119 and 120 whose experimental half-lives have not been found yet have been compared with the half-lives found from FRDM. The closed formula for the logarithm of half-life favorably explains the half-lives ranging from $10^{-6}$s to $10^{22}$y. Also this closed form expression curtains the dilemma over nonlinearity as it fairly reproduces the rectilinear alignment of the logarithm of the experimental decay half-lives as a function of parameter closely resembling with the Viola-Seaborg parameter. Thus having the half-life formula with us we can also predict the half-lives whose experimental values have not been found till date and these predictions will certainly give support to the experimentalists. \\

{\bf ACKNOWLEDGEMENTS:}

We gratefully acknowledge the computing and library facilities extended by the Institute of Physics, Bhubaneswar.


\begin{thebibliography}{99}
\bibitem{a} M. A. Preston, {\textit{Phys. Rev.}} {\bf 71}, 865 (1947).
\bibitem{b} I. Perlman, A. Ghiorso, and G. T. Seaborg, {\textit{Phys. Rev.}} 
	{\bf 77}, 26 (1950).
\bibitem{c} M. Balasubramaniam and N. Arunachalam, {\textit{Phys. Rev. C}} 
	{\bf 71}, 014603 (2005).
\bibitem{d} Y. Qian, Z. Ren, {\textit{ Phys. Lett. B}}  {\bf 738}, 87 (2014).
\bibitem{e} G. Gamow, {\textit{Z. Phys.}} {\bf51}, 204 (1928).
\bibitem{viola} V.Viola and G. Seaborg, {\textit{J. Inorg. Nucl. Chem.}} 
	{\bf 28}, 741 (1966).
\bibitem{patyk} A. Sobiczewski, Z. Patyk, and S. Cwiok, {\textit{Phys. Lett. B}}          {\bf 224}, 1 (1989).
\bibitem{pomorski} A. Sobiczewski and K. Pomorski, {\textit{Prog. Part. Nucl. Phys.}} {\bf 58}, 292 (2006).
\bibitem{ni} D. Ni, Z. Ren, T. Dong, and C. Xu, {\textit{Phys. Rev. C}} 
	{\bf 78}, 044310 (2008).
\bibitem{royera} G. Royer, {\textit{J. Phys. G: Nucl. Part. Phys.}} 
	{\bf 26}, 1149 (2000).
\bibitem{royerb} G. Royer, {\textit{ Nucl. Phys. A}} {\bf 848}, 279 (2010).
\bibitem{horoi} M. Horoi, {\textit{J. Phys. G: Nucl. Part. Phys.}} 
	{\bf 30}, 945 (2004).
\bibitem{poenaru} D. Poenaru, R. Gherghescu, and N. Carjan, {\textit{ Europhys. lett.}} {\bf 77}, 62001 (2007).
\bibitem{denisov} V. Y. Denisov and A. A. Khudebko, {\textit{At. Data Nucl. Data Tables}} {\bf 95}, 815 (2009).
\bibitem{sedykh} V. Y. Denisov, O. I. Davidovskaya, and I. Y. Sedykh, {\textit{Phys. Rev. C}} {\bf 92}, 014602 (2015).
\bibitem{shan} S. Zhang, Y. Zhang, Jianpo Cui, and Yanzhao Wang, {\textit{Phys. Rev. C}} {\bf 95}, 014311 (2017).
\bibitem{bsahu} B. Sahu and S. Bhoi, {\textit{Phys. Rev. C}} {\bf 93}, 044301 (2016).
\bibitem{bsahua} B. Sahu, R. Paira, and B. Rath, {\textit{Nucl. Phys. A}} 
	{\bf 908}, 40 (2013).
\bibitem{fur} V. I. Furman, S. Holan, S. G. Kadmensky, and G. Stratan,  
	{\textit{Nucl. Phys. A}} {\bf 226}, 131 (1974).
\bibitem{dav} C. N. Davids and H. Esbensen, {\textit{Phys. Rev. C}} {\bf C61}, 
	054302 (2000).
\bibitem{fro} C. E. Fr\"{o}berg, {\textit{Rev. Mod. Phys.}} {\bf 27}, 399 
	(1955).
\bibitem{fie} H. Fiedeldey, W. E. Frahn, {\textit{Annls.of Phys.}} {\bf16}, 
	387 (1961).
\bibitem{tieukuang} T. Dong and Z. Ren, {\textit{Eur. Phys. J.}} {\bf A26},
	69 (2005).
\bibitem{mohr} P. Mohr, {\textit{Phys. Rev. C}} {\bf C95}, 011302 (2017).
\bibitem{yu1} Yu. Ts. Oganessian et. al., {\textit{Phy. Rev. Lett.}} {\bf 74}, 
	044602 (2006).
\bibitem{yu2} Yu. Ts. Oganessian, {\textit{J. Phys. G: Nucl. Part. Phys.}} 
	{\bf 34}, R165 (2007).
\bibitem{frdm} P. Moller, {\textit{At. Data Nucl. Data Tables}} {\bf 66}, 131 
	(1997).
\bibitem{brog} R.A. Broglia and A. Winther. Heavy-ion reactions lecture notes. Addison-Wesley, Redwood City. 1981. p. 116.
\end{thebibliography}
\end{document}